\documentclass[3p]{elsarticle}

\journal{Experimental Mechanics}

\usepackage{amsmath}
\usepackage{xcolor}
\usepackage{soul}
\usepackage{mathtools}
\usepackage{setspace} 
\usepackage{float}

\usepackage{threeparttable}
\usepackage{etoolbox}
\usepackage{longtable}
\usepackage{multirow}
\usepackage{subfigure}    
\usepackage{booktabs}
\usepackage[utf8]{inputenc}
\usepackage{enumitem}
\usepackage{hyperref}

\definecolor{darkgreen}{rgb}{0.1,0.8,0.1}

\newcommand{\fig}{Figure~}
\newcommand{\insitu}{\textit{in-situ }}

\newcommand{\degree}{^{\circ}}

\newcommand{\ra}[1]{\renewcommand{\arraystretch}{#1}}

\begin{document}
\singlespace

\begin{frontmatter}

\title{A Nanomechanical Testing Framework Yielding Front\&Rear-sided, High-resolution, Microstructure-correlated SEM-DIC Strain Fields}

\author[mymainaddress]{T. Vermeij}
\address[mymainaddress]{Dept. of Mechanical Engineering, Eindhoven University of Technology, 5600MB Eindhoven, The Netherlands}
\author[mymainaddress]{J.A.C. Verstijnen}
\author[mymainaddress]{T.J.J. Ramirez y Cantador}
\author[secondaddress]{B. Blaysat}
\address[secondaddress]{Université Clermont Auvergne, Clermont Auvergne INP, CNRS, Institut Pascal, F-63000 Clermont-Ferrand, France}
\author[thirdaddress]{J. Neggers}
\address[thirdaddress]{Université Paris-Saclay, ENS Paris-Saclay, CentraleSupélec, CNRS, LMPS, 91190, Gif-sur-Yvette, France}
\author[mymainaddress]{J.P.M. Hoefnagels*}
\cortext[mycorrespondingauthor]{Corresponding author}
\ead{j.p.m.hoefnagels@tue.nl}

\begin{abstract}
The continuous development of new multiphase alloys with improved mechanical properties requires quantitative microstructure-resolved observation of the nanoscale deformation mechanisms at, e.g., multiphase interfaces. This calls for a combinatory approach beyond advanced testing methods such as microscale strain mapping on bulk material and micrometer sized deformation tests of single grains. We propose a nanomechanical testing framework that has been carefully designed to integrate several state-of-the-art testing and characterization methods, including: (i) well-defined nano-tensile testing of carefully selected and isolated multiphase specimens, (ii) front\&rear-sided SEM-EBSD microstructural characterization combined with front\&rear-sided \insitu SEM-DIC testing at very high resolution enabled by a recently developed InSn nano-DIC speckle pattern, (iii) optimized DIC strain mapping aided by application of SEM scanning artefact correction and DIC deconvolution for improved spatial resolution, (iv) a novel microstructure-to-strain alignment framework to deliver front\&rear-sided, nanoscale, microstructure-resolved strain fields, and (v) direct comparison of microstructure, strain and SEM-BSE damage maps in the deformed configuration. Demonstration on a micrometer-sized dual-phase steel specimen, containing an incompatible ferrite-martensite interface, shows how the nanoscale deformation mechanisms can be unraveled. Discrete lath-boundary-aligned martensite strain localizations transit over the interface into diffuse ferrite plasticity, revealed by the nanoscale front\&rear-sided microstructure-to-strain alignment and optimization of DIC correlations. The proposed testing and alignment framework yields front\&rear-sided aligned microstructure and strain fields providing 3D interpretation of the deformation and opening new opportunities for unprecedented validation of advanced multiphase simulations. DOI: \url{https://doi.org/10.1007/s11340-022-00884-0}

\end{abstract}


\end{frontmatter}

\section*{Highlights}

\begin{itemize}[labelindent=0.1em,labelsep=0.5cm,leftmargin=*]
\setlength\itemsep{0.08em}
\item Close integration of state-of-the-art nanomechanical testing, EBSD, and ultra-high resolution SEM-DIC
\item Extensive alignment framework of 8 independent data sets yields microstructure-correlated strain maps
\item Front\&rear-sided EBSD of microspecimens yields approx. microstructure over full 3D specimen volume
\item Forward-deformed strains \& microstructure mapped to post-mortem BSE damage maps, at both sides
\item Challenging test case: unraveling of nano-deformation mechanisms at martensite-ferrite phase boundary

\end{itemize}

\section{Introduction}

In materials science and engineering, the design and manufacturing of new materials and alloys is increasingly important to adhere to the ever more demanding requirements for materials, such as increased strength, ductility and toughness, for automotive and other industries. Many of these new materials, often metal alloys, have a multiphase microstructure in which, for instance, a hard phase is combined with a soft phase to retrieve favourable global properties that reflect a combination of the individual phases \cite{Rashid1981,pj2001developments, Mortensen2010,Tasan2015Review, CARROLL2015309, li2016metastable}. The further development of these materials is limited by our understanding of the governing deformation mechanisms at the micro- and nanoscale and the ability to cast this knowledge into numerical models for prediction of mechanical behaviour at the engineering scale. Such understanding can only be unlocked by means of detailed experimental analysis of the mechanics of these alloys. However, whereas almost all studies so far have focused on the contribution of the individual phases to the alloy's mechanics, recent works have shown that micro- and nanoscale deformation mechanisms at phase and grain boundaries may govern the material behaviour, in cases of early failure by interface damage \cite{Tasan2015Review, Landron2010, hoefnagels_dp_damage,LeiLiu2020} or increased ductility by deformation transfer over the phase interfaces and grain boundaries \cite{OROZCOCABALLERO2017367,WEI2021116520, harr2021effect}. Therefore, observation and quantification of plasticity (and subsequent damage) at these interfaces, at high resolution and \insitu during deformation, is crucial. Such studies are obviously challenging, but are often further complicated when one of the phases (such as martensite or bainite) has a fine and complex microstructure and a resulting complex (jagged) interface \cite{LeiLiu2020}, thereby leading to nanoscale microstructural and mechanistic features at the interfaces.

Through recent advances in experimental mechanics, (\textit{in-situ}) mechanical characterization of polycrystalline and multiphase metals at small scales is increasingly more accessible. Particularly, the development of Digital Image Correlation (DIC) patterning methods for \insitu Scanning Electron Microscopy based DIC (SEM-DIC) has lead to the capability of observing plastic deformation mechanisms from micrometer to nanometer scales \cite{kammers_daly_2011, di_gioacchino_quinta_da_fonseca_2012, YAN2015, Hoefnagels2019, montgomery2019robust, shafqat2021cool}. In practice, SEM-DIC testing of these metals is commonly performed on bulk samples that are (mechanically) polished, of which the microstructure is characterized (with, e.g., Electron Backscatter Diffraction (EBSD)), that are subsequently decorated with a DIC speckle pattern, and are finally tested by means of \insitu SEM-DIC, resulting in plastic strain fields over a certain millimeter or micrometer sized area in the microstructure \cite{OROZCOCABALLERO2017367, WEI2021116520, YAN2015, TASAN2014,TASAN2014386, STINVILLE201529, ZHANG201888, kumar2019situ, li2021interactions,Vermeij2021,charpagne2021multi}. Such approaches have provided valuable insights into plasticity mechanisms in the microstructure, strain partitioning between phases \cite{TASAN2014, TASAN2014386}, how plasticity leads to damage \cite{YAN2015, kumar2019situ,Vermeij2021}, and on qualitative deformation patterns at multiphase interfaces \cite{WEI2021116520, li2021interactions}. However, these experiments have severe limitations, as information on the subsurface 3D microstructure and on the true local stresses and boundary conditions are often unknown. Moreover, as said above, almost all studies focus on the deformation of the individual phases, whereas the quantitative interaction between the phases can strongly influence the individual response of the phases, thus requiring more detailed investigations. Aspects that complicate such experiments are (i) the inherent complexity and small scale of the microstructure, especially when complex phases such as martensite and bainite are involved, leading to a highly complex, intangible combination of deformation mechanisms near and at the interfaces and (ii) the inherent nanoscale dimensions of these microstructures and their mechanisms (e.g. individual slip traces and substructure sliding \cite{DU2016Plasticity}), requiring nanoscale spatial strain resolution and high-resolution microstructure to strain alignments \cite{Vermeij2021}.

In contrast to these microscale SEM-DIC experiments on "bulk" materials, a different class of small-scale mechanical testing techniques exists that is particularly suitable for single phases and single grains. Well-defined micro-pillar compression, micro- or nanotensile testing, and micro-bending experiments offer unique insights into the stress-strain behaviour and slip system activity of individual phases, when performed on individual grains or bi-crystals \cite{DEHM2018248,GHASSEMIARMAKI2014, MOTZ20054269,KIENER2008580,DuTensileTest,Tian2020}. Fabrication of microscopic specimens by Focused Ion Beam milling (FIB), the use of dedicated nanometer and nano-force resolution testing rigs, and careful pre-test alignment yields deformation tests with a well-known stress-strain state and crystallography, while allowing \insitu and/or \textit{post-mortem} identification of individual slip systems or crack paths. However, multiphase, or even polycrystalline, specimens are rarely studied with these methods, with some exceptions for lamellar metals \cite{EKH2018272,edwards2019transverse}. Moreover, singular multiphase interfaces have yet to be tested on these scales, to the best of our knowledge. The main challenges in conducting these micro-testing techniques on polycrystalline and multiphase specimens, specifically when the focus lies on interface mechanics, include (i) the selection and extraction of specimens that contain (only) the most interesting feature (e.g. an individual interface) \cite{du2019lath}, (ii) measurement of the nanoscale deformation fields, with high-resolution SEM-DIC, during these micro-tests, which is currently only achieved by a small number of research groups \cite{edwards2019transverse, di2014mapping, edwards2022mapping} and (iii) attribution of deformations to microstructure features and subsequent identification of their character, which requires data collection of two (or more) sides of the specimen and careful data alignment.

In this work, we present a new nanomechanical testing framework that addresses all these challenges as this framework allows to combine, for the first time, such a large number of state-of-the-art micro-mechanical testing and microscopic characterization tools to enable highly detailed investigation of nanoscale deformation mechanisms. These tools include (i) nano-tensile testing of "1D" ($3.5*2.5*10$\textmu m) specimens \cite{DuTensileTest}, isolated from the bulk microstructure at specific regions of interest, (ii) multi-modal microstructure measurements at the front and rear of the micro-specimens, (iii) front\&rear-sided SEM-DIC yielding very high-resolution strain maps by (iv) application of a recently proposed nanoscale DIC speckle patterning technique \cite{Hoefnagels2019,Vermeij2021}, (v) SEM scanning artefact correction \cite{neggerscancor} and (vi) DIC deconvolution correction \cite{grediac2019robust}, while (vii) all of this front\&rear-sided data is aligned using a novel data alignment framework. We will demonstrate that the combination of all these measurement modalities will yield front\&rear-sided nanoscale and microstructure-resolved strain fields that push the limits of SEM-DIC. A particularly interesting type of specimen is investigated in which an incompatible (i.e. high crystallographically misoriented) ferrite-martensite interface reveals a sharp transition from discrete (martensite) to diffuse (ferrite) plasticity. It will be shown that the nanoscale spatial strain resolution and the alignment framework are crucial to determine how the (near-)interface behaviour evolves, while the correlated front\&rear-sided microstructure and strain fields allow interpretation in 3D of both microstructure and deformation. A parameter study that includes the DIC subset size, DIC deconvolution and strain field calculation method reveals how both discrete (martensite) and diffuse (ferrite) plasticity can be identified, which is crucial for complex multiphase specimens. This specimen, and a second specimen produced in the same way, will be used to explain the nanomechanical testing framework in Section \ref{sec:framework}. Finally, the aligned data allows assessment of the deformation in the (amplified) forward deformed configuration, which eases interpretation and analysis of the micro-mechanics at and near the multiphase interfaces.

In Section \ref{sec:framework}, the details of the nanomechanical testing framework will be provided based on its 4 pillars: (I) specimen selection and fabrication, (II) characterization and nano-DIC patterning, (III) nanoscale testing and DIC, and (IV) data alignment. Subsequently, in Section \ref{sec:casestudy}, the capabilities of the framework will be demonstrated on a nano-tensile specimen containing a single and clean martensite-ferrite interface. To this end, a commercial DP600 grade (0.092C-1.68Mn-0.24Si-0.57Cr wt.\%) has been heat-treated (20 minutes austenization at 1000°C, followed by a cool down to 770°C in 50 minutes, 30 minutes inter-critical anneal at 770°C, and water quenching to room temperature) to produce a coarse ferrite-martensite microstructure with martensite volume fraction of $70 \pm 5 \%$ that allows for selection and fabrication of nano-tensile specimens with a single, continuous and straight ferrite-martensite interface that runs from the top surface to the bottom surface. Finally, discussion and conclusions follow in Section \ref{sec:conclusions}.

\section{Nanomechanical Testing Framework}
\label{sec:framework}

\subsection{Methodology Part I: Specimen Selection and Fabrication}

The specimen fabrication and characterization methods are based on the work of Du \textit{et al.} \cite{DuTensileTest} and are optimized and automated further in this work. \fig \ref{fig:Fig1}a-d shows the different steps of the sample preparation, specimen selection and specimen fabrication. A $12 \times 9 \times 1\ $ mm piece of material is mechanically ground and electropolished to create a 4° wedge with a deformation-free, high-quality surface at both sides of the micrometer-thin tip of the wedge (\fig \ref{fig:Fig1}a) \cite{DuTensileTest}.
 
\fig \ref{fig:Fig1}b illustrates the three-sided SEM characterization of the wedge tip (here done using a Tescan Mira 3) for identification of viable specimen locations on the $6\ $ mm long section of the wedge tip, which is highlighted in \fig \ref{fig:Fig1}a by the red rectangle and in \fig \ref{fig:Fig1}b by the yellow dashed lines. Long phase boundaries, perpendicular to the wedge tip, are first identified on the rear surface using channelling contrast in SEM BackScattered Electron (BSE) images. Then, SEM Secondary Electron (SE) images of the tip are used to spatially align the front and rear microstructure, after which a viable specimen location can be determined, see the red rectangle in \fig \ref{fig:Fig1}b.

The FIB milling procedure (done here with a FEI Nova Nano FIB-SEM Dual Beam) for specimens fabrication on the wedge tip is schematically shown in \fig \ref{fig:Fig1}c, where we employ the same FIB milling procedure as Du \textit{et al.} \cite{DuTensileTest}, with an added final parallel FIB polishing step for improved EBSD quality. Initially, a wider nano-tensile specimen is fabricated, centered around the identified location. The reason for this is that during the FIB thinning on the front side of the wedge tip to create a parallel front and rear surface, the material removal causes the frontside to change for a "non-columnar microstructure", which prevents precise identification of the final specimen location before FIB thinning. This problem is addressed by leaving a wider (parallel) specimen after the frontside thinning, from which the final, narrower specimen location is selected using BSE (and if necessary EBSD) imaging. For an optimal configuration of the ferrite-martensite boundary/interface, since the martensite is stronger than the ferrite, it is crucial to obtain the phase boundary along the full length of the specimen, in order to obtain "parallel" loading of both the ferrite and martensite phase, and to prevent early necking over a full ferrite cross-section \cite{Du2019}. The specimen is finished with a fine parallel FIB milling step. A finished specimen (\fig \ref{fig:Fig1}d) typically has a gauge section of $10 \times 3.5 \times 2.5\ $\textmu m $(Length \times Width \times Thickness)$, however dimensions may vary in order to obtain the desired microstructure.

\subsection{Methodology Part II: Characterization and Nano-DIC Patterning}
\begin{figure}[H]
	\centering
    \includegraphics[width=0.8\textwidth]{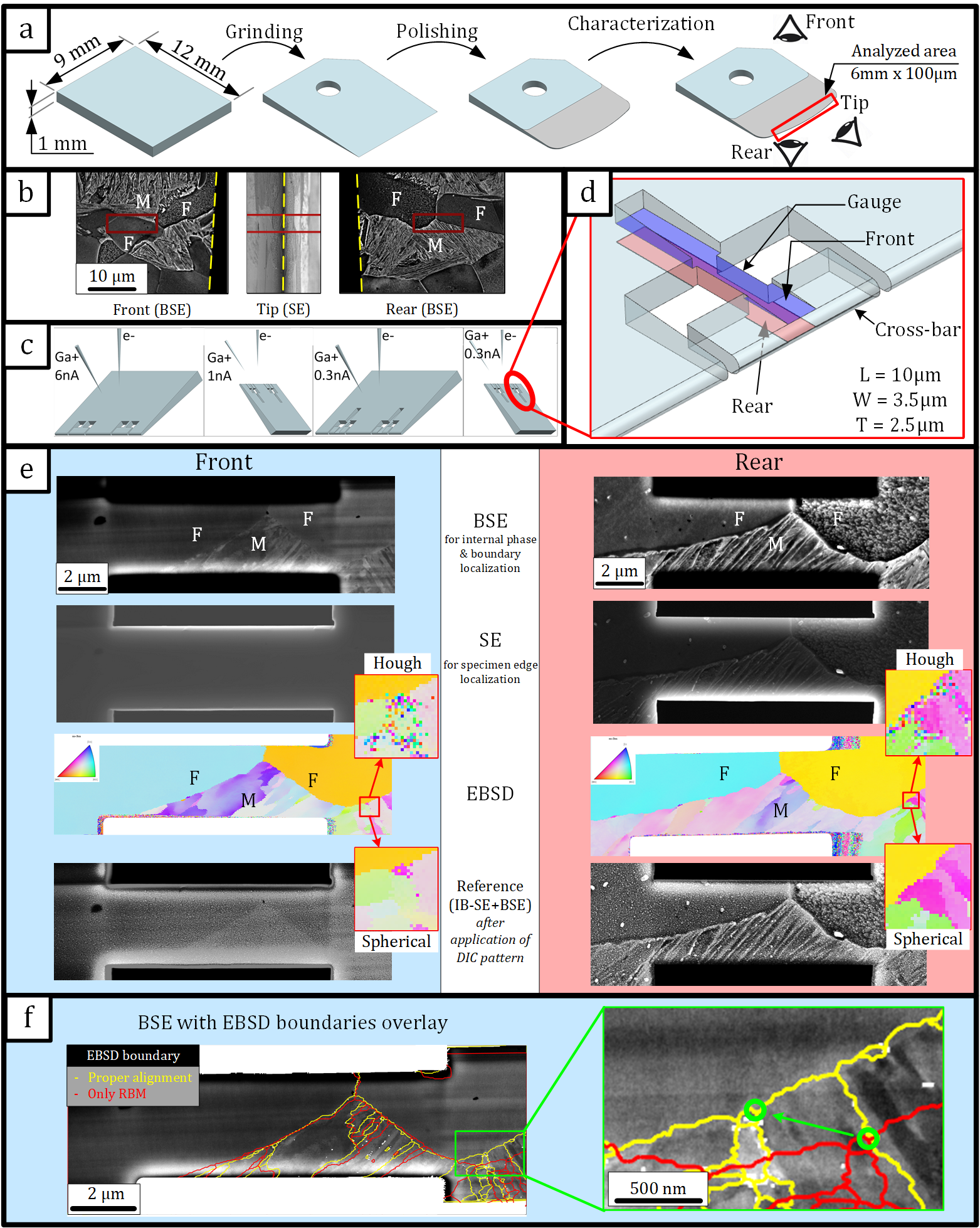}
    \caption{\small \textit{General overview of the sample preparation and characterization. (a) Wedge production process, detailing mechanical grinding and electropolishing for a deformation free, high quality, wedge tip. (b) Secondary Electron (SE) and BackScattered Electron (BSE) SEM images, used to identify possible specimen locations. The yellow dashed line indicates the wedge edge and the red areas indicates a viable specimen position. Viewing directions are indicated on the finished wedge in (a). (c) Schematic overview of Focused Ion Beam (FIB) milling steps used for specimen fabrication. From left to right: rough contouring, rough thinning, fine contouring, fine thinning \cite{DuTensileTest}. (d) A schematic of a finished specimen and its general dimensions. (blue = front, red = rear). (e) front\&rear-sided microstructure and EBSD crystallographic orientations (normal direction inverse pole figure) of a finished specimen, with insets showing the quality increase owing to Spherical Indexing with EMSphInx \cite{LENTHE2019112841} compared to standard indexing. All rear maps are flipped over the horizontal axis to allow easy comparison to the front. The reference image shows an overlay of in-beam SE (IB-SE) and BSE images after the DIC pattern is applied. (f) Experimental data with proper alignment (yellow) and without proper alignment (red, only rigid body movement). The green circles in the zoom annotate the same triple junction of the EBSD data.}}
    \label{fig:Fig1}
\end{figure}

SEM characterization is performed on the front and rear surfaces (colored blue and red respectively, in \fig \ref{fig:Fig1}) of the finished nano-tensile specimen. After FIB thinning, the microstructures on both sides of the specimen are more similar, but never 100\% the same. Therefore, microstructure information from both sides of the specimen is needed to estimate the full 3D microstructure for interpretation of the through-thickness plastic behavior and for potential comparisons to 3D simulations.

\fig \ref{fig:Fig1}e) shows BSE images which are used to determine the arrangement of phases in the microstructure and the morphology of the F/M interface, requiring optimization of SEM parameters for electron channeling contrast imaging \cite{LeiLiu2020}. While electropolishing induces a surface roughness that results in BSE edge effects and decreases the channeling contrast quality, especially at the phase boundaries, FIB milling on the front surface removes this topography and thereby allows more accurate identification of the location of the interface. However, since BSE electrons originate from a larger interaction volume, deeper inside the material, these images are ill-suited for identification of topography and of the exact location of the gauge edges and corners. Such information is required for (i) determination of the specimen geometry for calculation of global stress levels during the experiment and (ii) for several alignment steps in Section \ref{sec:Section_data_alignment_framework}. Therefore, SE images (\fig \ref{fig:Fig1}e) are captured, during the same scan as BSE, for assured data alignment, with optimized contrast and brightness settings for localization of the gauge edges and corners. Additionally, the specimen thickness is measured on SE images taken from the side and under an angle (not shown here). SE and BSE images acquired directly after FIB milling will, in the rest of this work, be referred to as the \textbf{Microstructure} dataset. 

EBSD measurements are performed (with an Edax Digiview 2 camera) on both front and rear sides of the specimen, with the raw EBSD patterns saved for subsequent offline indexing. The spherical cross-correlation indexing algorithm "EMSphInx" \cite{LENTHE2019112841} is employed for robust indexing of noisy (martensite) EBSPs, improving indexing of martensite and F/M interfaces considerably, as compared to traditional Hough transform based indexing. The Normal Direction Inverse Pole \fig (ND IPF) of specimen S1 is shown in \fig \ref{fig:Fig1}e, with insets highlighting the difference in quality between Hough indexing and Spherical indexing. This improvement of EBSD quality will turn out to be crucial for the alignment steps based on microstructural features. As will be demonstrated in future work, the resulting EBSD map of the crystallographic orientations can be used for quantification of slip systems, orientation relationships and plastic compatibility \cite{vermeij2022crystallographic,Vermeij2022boundarysliding,Vermeij2022dpcompatibility}. Furthermore, Confidence Index (CI) and Image Quality (IQ) are EBSD quality metrics that can be used for identification of lath boundaries and phases, as lath martensite has significantly higher dislocation densities than ferrite, thereby producing lower quality EBSD patterns \cite{Wright2006}. In this work, the crystallographic orientation, CI and IQ maps will be referred to as the \textbf{EBSD} dataset.

Subsequently, to enable strain measurements at the nanoscale, a scaleable and dense nano-DIC speckle pattern was applied using a recently developed novel patterning method, involving a single-step sputter deposition of a low temperature solder alloy that forms nanoscale islands during deposition \cite{Hoefnagels2019, Vermeij2021}. The solder alloy In52Sn48 was sputtered at $18\ $mA, $1E-2\ $mbar, 25°C, $120\ $s, at $\sim80\ $mm from the sputter target, resulting in a dense, random, high-quality pattern with features of $\sim20-50\ $nm in size as shown in \fig \ref{fig:fig2}a. The patterning parameters deviate slightly from (c) in Table 1 of Hoefnagels \textit{et al.} \cite{Hoefnagels2019}, in order to acquire a slightly finer nano-DIC pattern ($20-50\ $nm vs $20-100\ $nm), which also reduce the InSn thickness for improved BSE imaging. For the first time in literature, this nano-DIC pattern is applied to both the front and rear of the nano-tensile specimen surfaces, in separate patterning steps, allowing for front\&rear-sided SEM-DIC, albeit that for the rear side SEM-DIC can only be applied on the images captured before and after the test (not \textit{in-situ}). 

After nano-DIC patterning, the specimens are again subjected to SEM characterization with high acceleration voltage ($20\ $keV) for correlation of the DIC pattern with the exact underlying microstructure. The high acceleration voltage creates an interaction volume which penetrates through the nanometer thickness DIC speckle pattern, allowing for in-beam BSE (IB-BSE) and regular BSE imaging of the underlying microstructure \cite{Vermeij2021}. Despite the large interaction volume, in-beam SE (IB-SE) and regular SE images still provide representative images of the nano-DIC speckle pattern as these signals mainly originate from the top surface. The high voltage (IB-)SE and (IB-)BSE images, acquired after application of the DIC speckle pattern, will, for the rest of this work, be referred to as the \textbf{Reference} dataset. This dataset is used as the baseline onto which all other data will be warped and aligned, as it can be easily correlated to all other datasets. An overlay of IB-SE and BSE data is provided in the bottom row of \fig \ref{fig:Fig1}e, illustrating the correlation between speckle pattern and underlying microstructure. 

\subsection{Methodology Part III: Nanoscale Testing and DIC}
\label{sec:dicmethod}

For \insitu mechanical loading of the microscopic specimens, many types of testing setups can be used, such as an \insitu nano-indentation stage for micro-compression or micro-bending tests (e.g. using a Hysitron Picoindenter). Requirements for the stage include: (i) the capability for \insitu SEM at low acceleration voltages, and (ii) the ability to mount the wedge, and to load the specimens that are fabricated on the tip, to enable precise (axial) loading under one-sided \insitu SEM observation. Additionally, the rear (opposite the \insitu measured) surface nano-DIC pattern requires measurement \textit{ex-situ}, i.e. before and after performing the \insitu test. 

In this work, the nano-tensile testing procedure of Refs. \cite{DuTensileTest, Bergers2014} has been extended to \insitu SEM-DIC by integrating the Nano-Tensile Stage (NTS) into  the SEM (in this case a Tescan Mira 3). In addition, a procedure for sequential \insitu SEM-DIC testing of multiple nano-tensile specimens (located on the same wedge), without recurring optical alignment, was developed. All nano-tensile specimens on a single wedge are produced parallel to the rear surface, however, as the wedge tip may not be perfectly straight and specimens on the wedge may be produced during multiple FIB sessions, small relative misorientations may be present between the specimens, which must be known to allow sequential \insitu SEM-DIC testing of multiple specimens without breaking vacuum. Therefore, before the wedge is mounted in the NTS for mechanical testing, optical profilometry and SEM imaging are employed to measure, respectively, the relative tilt and relative in-plane misorientation of all specimens. The wedge is then mounted in the NTS and a single specimen is aligned, along the two rotational axes that are perpendicular to the loading axis, using optical microscopy and optical profilometry, following the procedure of Bergers \textit{et al.} \cite{Bergers2014} and Du \textit{et al.} \cite{DuTensileTest}, yielding near perfect uniaxial tension alignment, resulting in less than 0.5\% bending stress at the start of the test. After a tensile test has been finished, the next specimen can be aligned inside the SEM with similar accuracy based on the previously measured relative misorientation with respect to the previous specimen. All specimens can be tested individually without affecting the others, since they are sufficiently separated on the wedge tip.

\begin{figure}[H]
	\centering
    \includegraphics[width=0.8\textwidth]{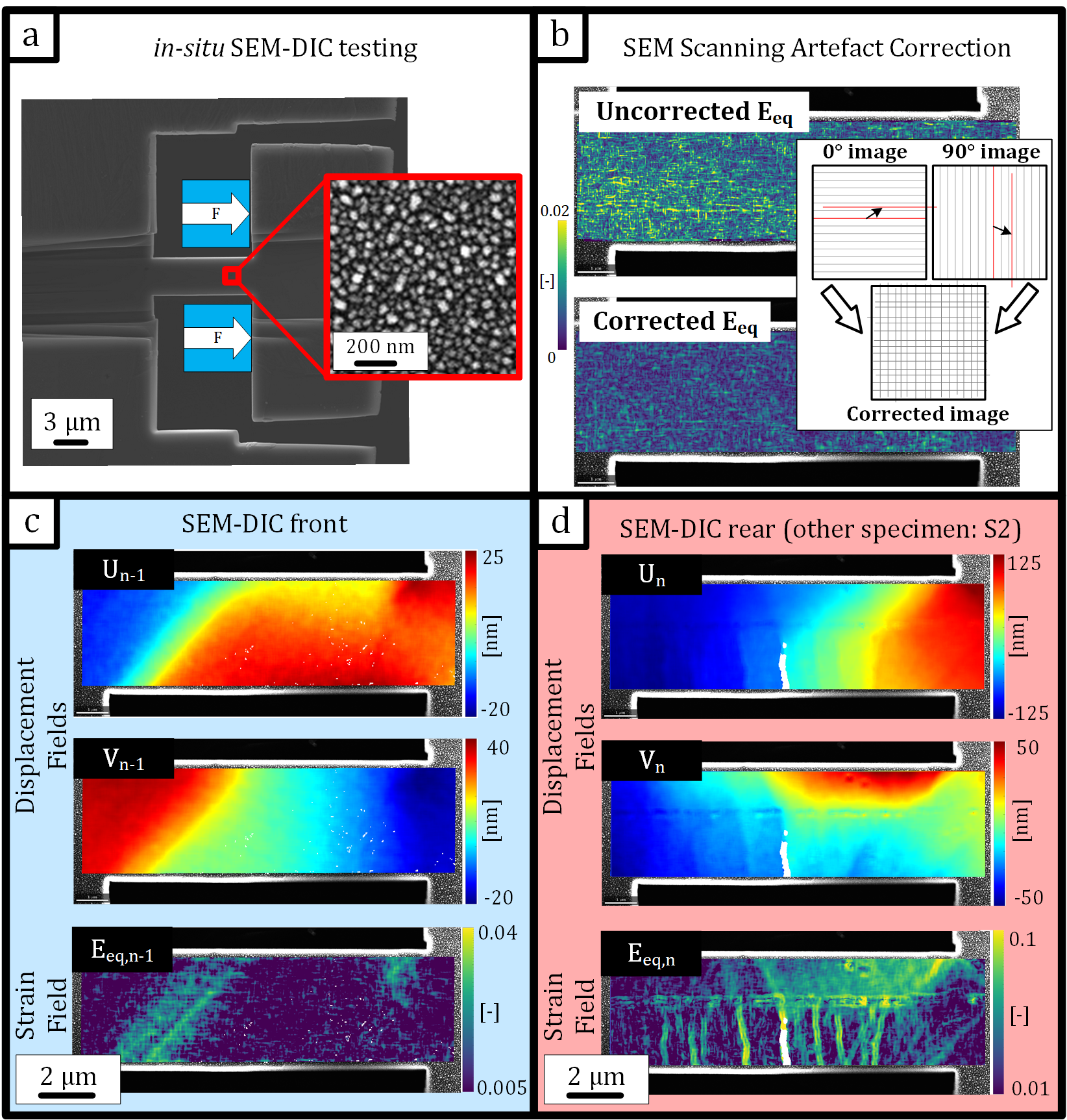}
    \caption{\small \textit{SEM-DIC nano-tensile testing. (a) Example of a specimen with an inset showing the nanoscale DIC pattern. The blue squares in the left image represent the gripper "arms" with specified loading direction that is precisely aligned to the specimen axis resulting in clean uniaxial tension \cite{DuTensileTest, Bergers2014}. (b) Deformation-free equivalent strain fields without and with scanning artefact correction applied \cite{neggerscancor}, with a schematic drawing of the correction to illustrate the concept of the correction method. (c) Front side SEM-DIC displacement and equivalent strain fields for increment $n-1$, i.e. second-to-last increment, of specimen S1 and (d) rear side SEM-DIC displacement and equivalent strain fields for the final (unloaded and \textit{post-mortem}) increment $n$ of specimen S2.}}
    \label{fig:fig2}
\end{figure}

\textit{In-situ} SEM-DIC uniaxial nano-tensile experiments (\fig \ref{fig:fig2}a) are performed on the aligned specimens. At pre-determined global stress levels, the loading is paused for IB-SE and SE imaging of the front surface with a low acceleration voltage ($5\ $kV) at a low working distance ($3.5\ $mm), limiting the interaction volume as much as possible for optimal contrast and spatial resolution of the nano-DIC pattern (inset of \fig \ref{fig:fig2}a). Note that this low working distance, which was chosen after careful optimization of the nano-DIC pattern imaging resolution and quality, is not a hard requirement for the method. Two images with orthogonal scanning directions (by applying $90\degree$ scan rotation for the second scan) are acquired at each stress level and are "combined", using a recently developed tool to correct for SEM scanning artefacts, i.e. the so-called \textbf{"Scan\_Corr"} GUI developed by Neggers \textit{et al.} \cite{neggerscancor, Teyssedre2011}. This correction algorithm assigns translation degrees of freedom to each scan line of the two orthogonal scans in order to find the best match between the scans in a dedicated DIC framework, whereafter these two adjusted images are combined into a single corrected image. This is based on the assumption that the fast scanning direction, i.e. the scan lines, in SEM images contain considerably fewer artefacts than the slow scanning direction, wherein drift and line jumps are prone to distort the SEM image \cite{sutton2007scanning, maraghechi2019correction}. \fig \ref{fig:fig2}b shows a deformation-free strain field without and with SEM artefact correction applied, supplemented by a schematic overview of the correction method. The strain field after correction shows a decrease in the overall noise level and, more importantly, fewer systematic line artefacts, which could be mistaken as e.g. slip activity. These line artefacts were likely introduced by line jumps and minor vibrations during testing. Further details and examples of the SEM artefact correction method are described by Neggers \textit{et. al.} in Ref. \cite{neggerscancor}. An overview of the different SEM and EBSD imaging parameters are provided in Table \ref{Table:SEM_params}.

During the \insitu test, we aim to acquire images at several increments with increasing plastic deformation, without fracture of the specimen, as imaging of the deformed rear surface is only possible after un-mounting the wedge from the tensile stage, here called the \textbf{Post-mortem} dataset. By avoiding fracture, SEM-DIC can be performed on the full rear surface (for the last deformation step), providing valuable information regarding the presence or absence of continuity of plasticity through the thickness. Local DIC is employed to correlate the artefact-corrected SEM-DIC images to retrieve the displacement field $\vec{u}$. For noise reduction, a gaussian filter with a standard deviation of 3 datapoints is applied to the displacement data. Thereafter, the displacement gradient tensor $\vec{\nabla}_0\vec{u} $ is calculated by taking the direct (nearest-neighbours) spatial gradient of $\vec{u}$ and is used to compute the Green-Lagrange strain tensor: $\mathbf{E}=\frac{1}{2}[(\vec{\nabla}_0\vec{u})^T+\vec{\nabla}_0\vec{u}+(\vec{\nabla}_0\vec{u})^{T} \cdot  \vec{\nabla}_0\vec{u}]$. To simplify the analysis and the plotting of local strains, we employ the 2D equivalent von Mises strain measure, which has been shown to be a reliable indicator for local plasticity \cite{TASAN2014386, Vermeij2021}: $E_{eq}=\frac{\sqrt{2}}{3}\sqrt{(E_{xx}-E_{yy})^2+E_{xx}^2+E_{yy}^2+6E_{xy}^2}$. The DIC reference image and displacement/strain fields as described here, are, for the rest of this section, referred to as the \textbf{SEM-DIC} dataset. In this section all artefact-corrected SEM-DIC images have been correlated using the commercial DIC package "MatchID", using a pixel size of $7\ $nm, subset size of 29 pixels and a step size of 1 pixel (see also Table \ref{Table:DIC_params}). In the case study in Section \ref{sec:casestudy}, we will study the influence of a different subset size and a different strain calculation method, employing a home-built DIC code that can be executed with and without applying so-called "deconvolution" \cite{grediac2019robust}, to demonstrate (i) the high spatial resolution of the strain field that can be achieved and (ii) the capability to identify both diffuse and discrete slip through different DIC strategies.

\fig \ref{fig:fig2}c shows an example of the resulting SEM-DIC displacement (U and V) and equivalent strain fields ($E_{eq}$) of specimen S1 for the second-to-last increment, $n-1$, where this specimen has not fractured yet. Several slip bands can be discerned in both displacement and strain fields. Since S1 did fracture, specimen S2 is used to illustrate the rear displacement and strain fields in \fig \ref{fig:fig2}d. In the case study of S2 in Section \ref{sec:casestudy}, we will showcase how the front and rear strain fields can be interpreted for a full 3D view of the deformation mechanisms. It is clear from the strain fields in \fig \ref{fig:fig2}c-d that we need a proper alignment from the \textbf{Microstructure}, \textbf{EBSD} and \textbf{SEM-DIC} data sets to the \textbf{Reference}, to be able to relate the microstructure-resolved strain fields to the (complex) nanoscale deformation mechanisms.

\begin{table}[H]\centering
\caption{\textit{\small SEM parameters used during acquisition of the \textbf{Microstructure}, \textbf{EBSD}, \textbf{Reference}, \textbf{SEM-DIC} and \textbf{Post-mortem} data sets (HV = high voltage, i.e. the SEM acceleration voltage, BI = beam intensity (a Tescan beam current measure), WD = working distance, FOV = field of view, BW = bandwidth for spherical indexing using EMSphInx \cite{LENTHE2019112841}). All data is captured with a Tescan Mira 3; when SEMs from other manufactures would be used, these settings may have to be adjusted slightly.}}
\label{Table:SEM_params}
\ra{1.3}
\begin{tabular}{@{}lcccccr@{}}\toprule
\textbf{Data set:} & \textbf{Microstructure} & \textbf{EBSD}&\textbf{Reference} & \textbf{SEM-DIC} & \textbf{Post-Mortem} &\textbf{Unit}\\
\midrule
Detector        & BSE & EBSD & (IB-)SE+BSE & IB-SE+SE & BSE &-   \\ 
SEM Mode        & Depth & Depth & Depth & Resolution & Depth    &   -\\ 
HV    		& 20    & 20 	& 20    & 5      &20       & kV           \\ 
BI    		& 18    & 18 	& 16    & 7          &16   &  -          \\ 
WD          & 10.5  & 20 	& 8.8   & 3.5  & 8.8           & mm \\ 
Tilt		& 0	& 70	& 0	& 0 &0				& °\\
Dwell time  & 32    & 8000 & 32    & 8      &32       & \textmu s/pix  \\ 
FOV         & 15    & - 	& 21.5  & 21.5      &21.5    & \textmu m   \\ 
Resolution  & 2048$\times$2048     & - & 3072$\times$3072      & 3072$\times$3072      & 3072$\times$3072 & pix $\times$ pix \\ 
Binning	& -	& 8$\times$8& -	& - &- & pix $\times$ pix\\
pixel size 	& 7.3 	& 30	&	7	&			7 &7	& nm\\
BW (EMSphInx)& -	& 88	&	-	&		-		&-& -\\
\bottomrule
\end{tabular}
\end{table}

\begin{table}[H]\centering
\caption{\textit{\small Local DIC (MatchID) correlation parameters used in Section \ref{sec:framework}}}
\label{Table:DIC_params}
\ra{1.3}
\begin{tabular}{@{}lrl@{}}\toprule
\textbf{Parameter}  & \textbf{Value}    & \textbf{Unit} \\
\midrule
Software           & MatchID                &            \\ 
Image filtering           & Gaussian; std 0.67                & pix          \\ 
Subset size           & 29                & pix           \\ 
Step size             & 1                 & pix           \\ 
Matching criterion & ZNSSD & \\
Interpolant & Bicubic Spline & \\
Strain window & 3x3  & \\ 
Virtual Strain gauge size \cite{dicguide2018} & 41  & pix\\ 
Subset Shape function        & Affine            &               \\ 
\bottomrule
\end{tabular}
\end{table}

\subsection{Methodology Part IV: Alignment of All Front and Rear Data}
\label{sec:Section_data_alignment_framework}

Through the (\textit{in-situ}) experiments, several datasets are acquired, yet a precise relation between the coordinate systems is missing. In general, errors between two SEM datasets are only induced by rigid body motion (RBM), i.e. translation and (minor) rotation, whereas the \textbf{EBSD} dataset is often severely warped and has inaccuracies of the absolute crystal orientations \cite{NOLZE2007172}. Indeed, when the \textbf{EBSD} dataset is aligned to the \textbf{Microstructure} dataset using translations only, it is observed that the \textbf{EBSD} grain boundaries will never precisely correspond to the microstructure, as shown in \fig \ref{fig:Fig1}f with the red boundaries.

\begin{figure}[H]
	\centering
    \includegraphics[width=1\textwidth]{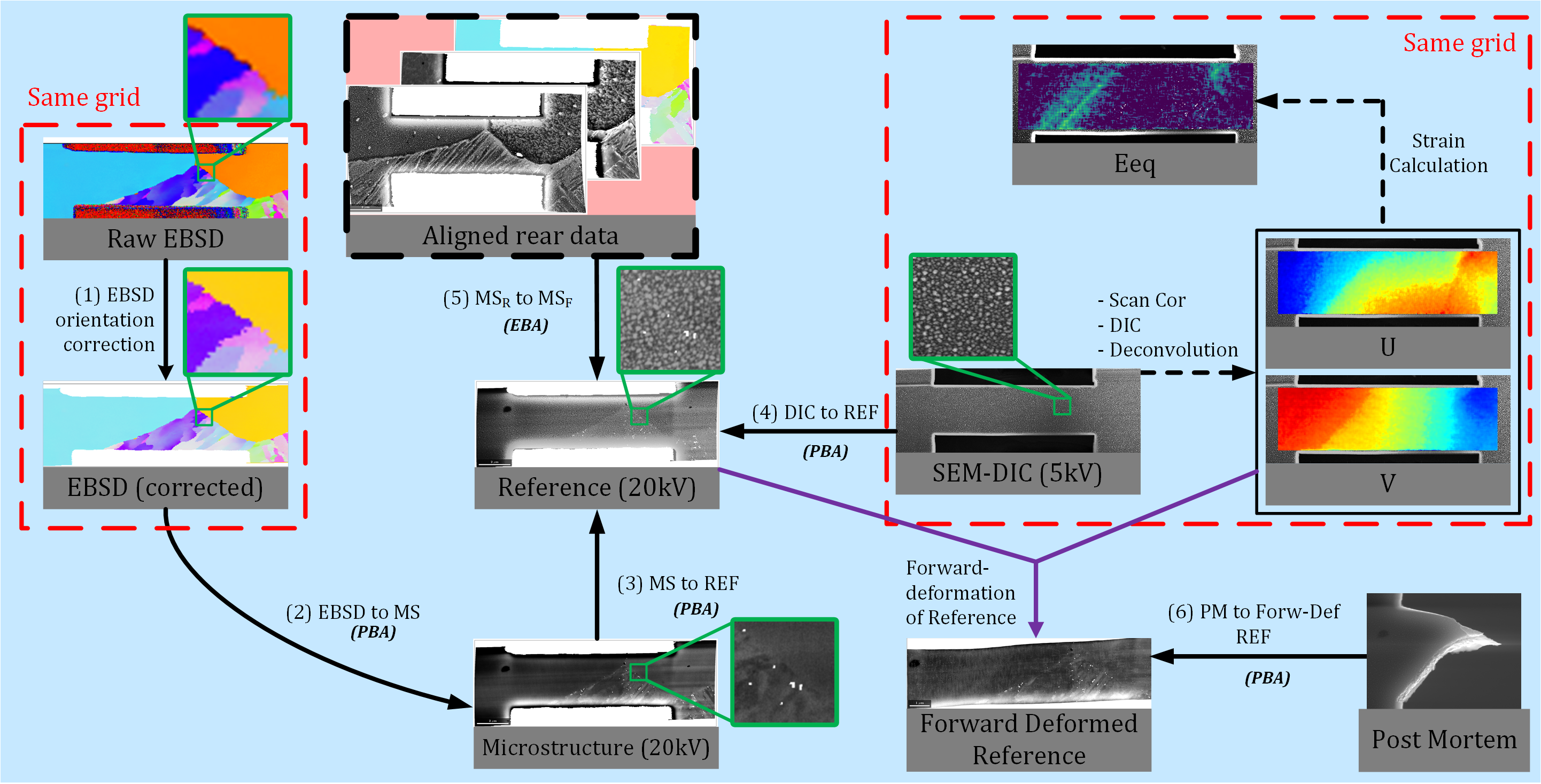}
    \caption{\small \textit{Schematic representation of all experimental datasets and alignment steps (EBSD = Electron backscatter diffraction, $MS_{F/R}$ = Microstructure (front/rear), REF = Reference, DIC = SEM-DIC, PM = Post-Mortem, EBSD orientation correction, PBA = Points-based alignment, EBA = edge-based alignment). Most datasets include a zoom (green border) of the same $1*1$ \textmu m area. Red borders indicate data manipulation without performing any spatial alignment resulting in altered data on the same grid. The alignment of the rear data is performed using the same steps (indicated by (1) to (4)) as shown in this figure for the frontside. The bottom-right of the figure also shows alignment in the deformed configuration, where a PM image is aligned to the forward-deformed REF configuration, which in turn is obtained by applying the DIC displacement field to each pixel (purple arrows).}}
    \label{fig:fig3}
\end{figure}

A novel data alignment framework is introduced, which is schematically shown in \fig \ref{fig:fig3}, that will reduce the misalignments between the datasets tremendously by not only correcting for translations and rotations, but also image warping, while simultaneously retrieving the correct absolute crystal orientations. Before the alignment framework with its different modules is explained, the improvements in alignment when using the framework is demonstrated for specimen S1 can be observed in \fig \ref{fig:Fig1}f (by comparing the red boundaries to the yellow ones). For emphasis, a single, arbitrary, triple junction is highlighted (\fig \ref{fig:Fig1}f, green circle) for both red and yellow boundaries, showing a decrease in misalignment of several hundreds of nanometers. For this specimen, the alignment framework made it possible to reduce the misalignment between the \textbf{EBSD} dataset and \textbf{Reference} dataset to 50 nm on average (close to the spatial resolution of EBSD), the \textbf{Microstructure} dataset and \textbf{Reference} dataset to below 50 nm, and the \textbf{SEM-DIC} datasets and the \textbf{Reference} dataset to as little as 15 nm. The framework also reduces the inaccuracies of the crystal orientations through an EBSD orientation correction step.

The data alignment framework is developed with Matlab using the MTex toolbox \cite{MTEX,bachmann2010} and is optimized for \insitu SEM-DIC nano-tensile experiments of specimens with a heterogeneous microstructure. Three self-contained modules are employed for alignment and correction (\fig \ref{fig:fig3}): \textit{EBSD orientation correction}, \textit{points-based alignment} (PBA) and \textit{edge-based alignment} (EBA). This modular design allows for application to the front and rear datasets, providing flexibility in alignment strategy, also for other types of experiments (e.g. single-sided SEM-DIC tests, e.g. on bulk material). In the upcoming description, the different alignment steps will be demonstrated for the more challenging front surface (of specimen S1), as FIB milling reduces the topographical contrast on this side of the specimen, complicating the alignment compared to the rear surface. Another important note is that successful alignment of the front surface is more valuable than alignment of the rear, as strain evolution can only be measured on the front surface during \insitu SEM-DIC nano-tensile experiments. Finally, as illustrated on the bottom of \fig \ref{fig:fig3}, we also perform a points-based alignment in the (final) deformed configuration, between the DIC-based forward-deformed \textbf{Reference} and the \textit{post-mortem} BSE images (taken at high eV to penetrate through the nano-DIC pattern). The Matlab code of the full alignment framework will be available on GitHub (\url{https://github.com/Tijmenvermeij/NanoMech_Alignment_Matlab}).

\subsubsection{EBSD Orientation Correction}

\begin{figure}[H]
	\centering
    \includegraphics[width=1\textwidth]{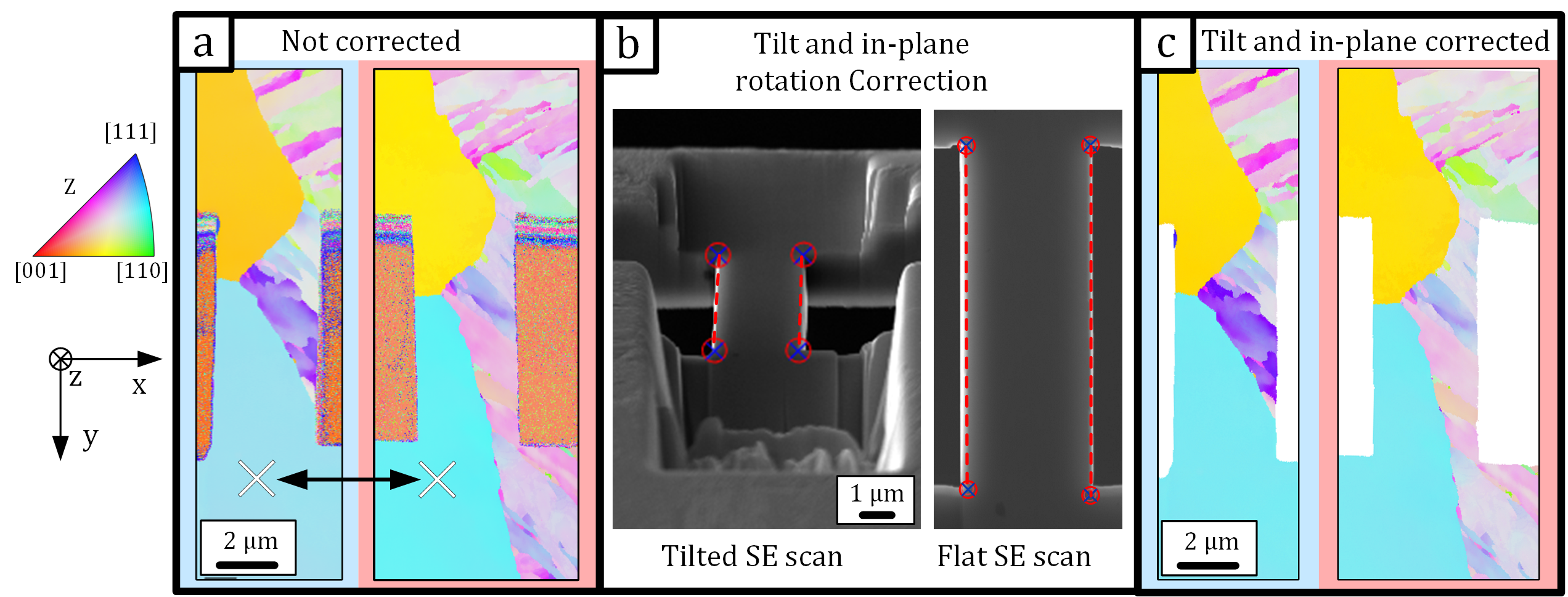}
    \caption{\small \textit{Overview of the EBSD orientation correction: (a) normal direction (ND) inverse pole figure (IPF) maps of the specimen, before correction, of the front and rear side. The white crosses indicates the same through-thickness grain, which has slightly different IPF colors on the front and rear side, (b) tilt and in-plane misalignment correction, by assessment of the length and gauge edge direction of the specimen, in tilted (left) and flat (right) configuration, with the markers showing homologous points and the red dotted lines showing the results of edge detection and fitting. (c) Front and rear ND IPF after all EBSD correction steps; note the same IPF colors of the front and rear ferrite grain, for both ferrite grains.}}
    \label{fig:fig4}
\end{figure}

In these experiments, the front\&rear-sided EBSD measurements allow us to assess the EBSD orientation accuracy by comparing the orientation of through thickness ferrite grains on front and rear sides. \fig \ref{fig:fig4}a shows the uncorrected EBSD maps of the front and rear surface (flipped over the vertical to allow easy comparison to the front), wherein the cyan and yellow ferrite grains appear to differ in their IPF color (e.g., as indicated by the white crosses). Indeed, the initial crystallographic misorientation between the front and rear \textbf{EBSD} dataset is measured to be, in so-called "axis-angle" representation \cite{diebel2006representing}, $4.1^o(-0.88\Vec{e_x }-0.08\Vec{e_y}+0.48\Vec{e_z})$. For other specimens, similar initial misalignments have been observed. Several sources of this misalignment can be identified, including the wedge and specimen production process (e.g. wedge curvature) as well as alignment errors (of, e.g., the wedge orientation and EBSD detector alignment) \cite{DuTensileTest,NOLZE2007172}. In this case, the rotation axis has large $\Vec{e}_x$ and $\Vec{e}_z$ components, indicating that most of the misalignment arises, respectively, from tilt and in-plane rotation. While the magnitude of absolute misorientation is mostly not being discussed in literature, it is clearly important to correct these to enable accurate identification of deformation mechanisms and slip activity. The \textit{EBSD orientation correction} module of the alignment framework employs 3 steps to reduce the error of the absolute crystallographic orientations: (a,b) for each side of the specimen separately, alignment of (a) tilt and (b) in-plane rotation to the tensile direction, by means of, respectively, points-based and edge-based alignment of the tilted SE scan at 70° to the flat SE scan at 0°, as shown in \fig \ref{fig:fig4}b, and (c) to determine the real crystallographic orientation, which is defined as the "averaged" orientation between the front and rear \textbf{EBSD} data. The mismatch in tilt and rotation are measured on SE images as these show sharp edges and corners, allowing for detection of corners (points) and edges, and have relatively low drift-induced artefacts, as compared to the EBSD data, as acquisition times are much lower.

For the "tilt correction" (a), the change in gauge length is determined between flat and tilted configuration, for the left and right gauge edges, by selecting the gauge's vertices, indicated with the red and blue markers in \fig \ref{fig:fig4}b. For specimen S1, the "true tilt" of the front and rear \textbf{EBSD} data was measured to be $66.7^o(\Vec{e}_x)$ and $72.4^o(\Vec{e}_x)$ respectively, resulting in tilt corrections of $3.3^o(\Vec{e}_x)$ and $-2.4^o(\Vec{e}_x)$ respectively. Next, the in-plane rotation mismatch (b) is found by measuring the misalignment of the specimen gauge edges, between tilted and flat SE scans. Canny edge detection and subsequent fitting of a line over the gauge edges (red dotted line in \fig \ref{fig:fig4}b) results in an average in-plane misalignment, which was $0.95^o(\Vec{e}_z)$ and $-1.43^o(\Vec{e}_z)$ for the front and rear of S1 respectively. Now, the front and rear \textbf{EBSD} data are corrected for these misalignments, using the tools available in MTex.

With the orientation errors in both front and rear \textbf{EBSD} data sets minimized separately, a final "combined" correction (c) of front and rear is performed. Here, the assumption is made that a single through-thickness ferrite grain has the same crystallographic orientation on both sides of the specimen. In practice, not all grains (e.g. in martensite) are identifiable through the thickness, making it impossible to perform this correction on all grains individually. Instead, the complete front and rear \textbf{EBSD} datasets are corrected based on the misorientation of a single large, through-thickness grain (indicated with the white cross in \fig \ref{fig:fig4}a) between front and rear. The required correction for specimen S1 is $2.1^o(0.15\Vec{e}_x+0.97\Vec{e}_y-0.20\Vec{e}_z)$ and is performed by applying the rotation from original towards averaged grain orientation, for front and rear \textbf{EBSD} data on the complete maps. This leads to an upper limit of the uncertainty (i.e. mismatch) on the crystallographic orientation of $\pm 1.05^o$, which is well below the generally accepted absolute EBSD accuracy of $\pm2^o$ \cite{NOLZE2007172}. The corrected ND IPFs of front and rear \textbf{EBSD} are shown in \fig \ref{fig:fig4}c, thresholded at $CI>0.2$.

\subsubsection{Alignment of the \textbf{EBSD} to \textbf{Microstructure} Datasets}

For all specimens in this work, direct alignment of the \textbf{EBSD} dataset to the \textbf{Reference} dataset is inhibited by the poor channeling contrast of small micro-structural features in the \textbf{Reference} (IB-)BSE images, which is caused by the InSn DIC speckle pattern obscuring the BSE signal and by the flat topography after FIB milling. Therefore, the \textbf{EBSD} data will first be warped, aligned and put on the same grid as the \textbf{Microstructure} data, which has better channelling contrast, using the \textit{points-based alignment} module of the alignment framework (step (2) in \fig \ref{fig:fig3}). Afterwards, The \textbf{Microstructure} (with \textbf{EBSD}) data will be aligned to the \textbf{Reference} data during step (3) "MS to REF", see \fig \ref{fig:fig3}.

\begin{figure}[H]
	\centering
    \includegraphics[width=0.8\textwidth]{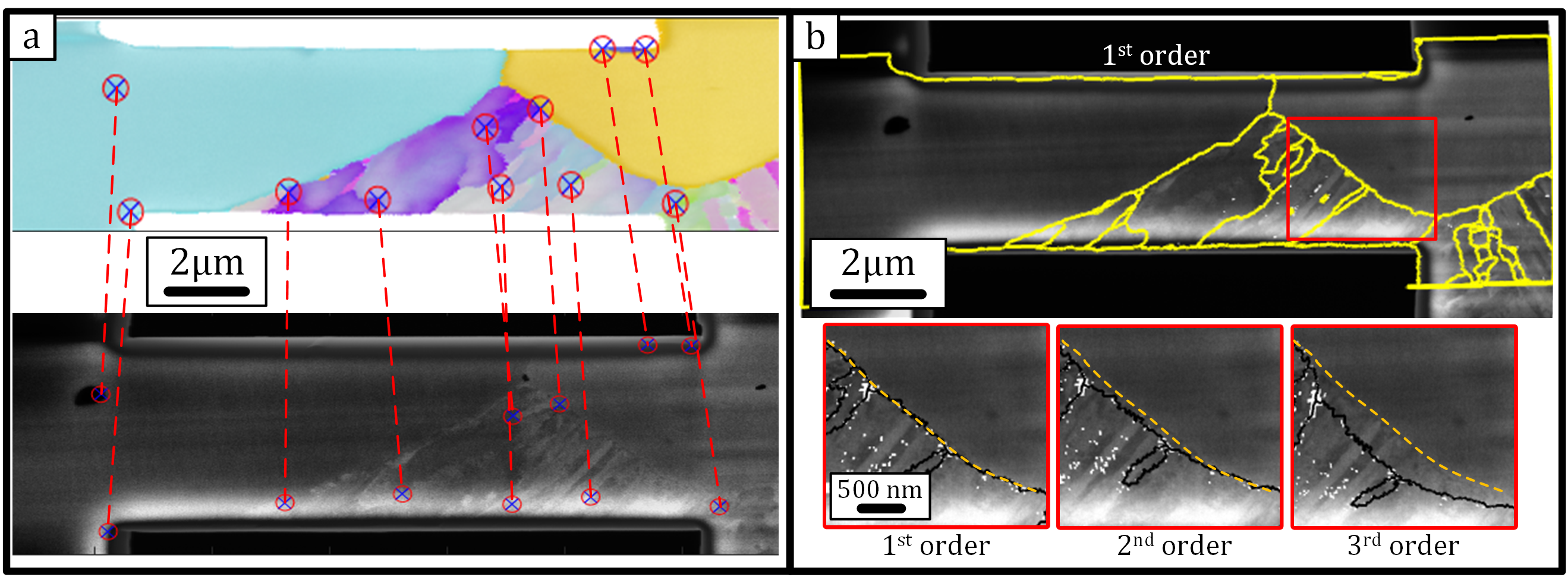}
    \caption{\small \textit{Alignment of \textbf{EBSD} to \textbf{Microstructure} data sets, demonstrated for specimen S1. (a) shows the \textbf{EBSD} ND IPF with 50\% transparent CI overlay and the SE/BSE overlay data from the \textbf{Microstructure} dataset, with selected homologous points in both datasets. The dashed red lines connect corresponding selection points. (b) \textbf{Microstructure} BSE/SE overlay with aligned (and warped) \textbf{EBSD} grain boundaries plotted as overlay. A small area with a clear F/M interface (manually drawn over the BSE image with an orange dashed line) shows the alignment results (aligned grain boundaries in black lines) for 1st, 2nd and 3rd order polynomial displacement fields, using the same set of homologous selection points.}}
    \label{fig:fig5}
\end{figure}

Several distinct points can be identified in the \textbf{EBSD} ND IPF with CI overlay and \textbf{Microstructure} SE/BSE overlay in \fig \ref{fig:fig5}a, in which pairs of homologous selection points, i.e. features on the specimen that can be identified in both datasets, are connected with dashed red lines. The connection lines are not parallel, nor of equal length, indicating that the \textbf{EBSD} data is severely warped with respect to the \textbf{Microstructure} data. 

For all pairs of selection points the misalignments are calculated ($\Vec{U} = \Vec{x}_{REF}-\Vec{x}_{EBSD}$) and used to fit a polynomial displacement field needed to align the \textbf{EBSD} data to the \textbf{Microstructure} data. The polynomial order, individually determined for each alignment step, should be chosen as low as possible, as unnecessary high polynomial orders may introduce large errors \cite{ATKINSON2020110561}. Alignment and warping of the \textbf{EBSD} data to the \textbf{Microstructure} data is performed using the polynomial displacement fields:  
\begin{equation}
    \label{eq:affine_disp}
    \begin{cases}
        U_x = \displaystyle\sum_{k=0}^{n} a_k x^i y^j,\\
        U_y = \displaystyle\sum_{k=0}^{n} b_k x^i y^j,\\
        i+j \leq k,
    \end{cases}
\end{equation}
with $n$ constituting the polynomial order. For a total of N degrees of freedom (DOFs), e.g. $N=6$ in the case of a first order polynomial, $\frac{N}{2}$ selection points are required, at a minimum, to (uniquely) describe the polynomial displacement fields, as each selection point yields both an x and y coordinate. The use of more selection points is recommended, as it results in least-squares fitting of the polynomial field, thereby minimizing errors that result from erroneous selection points. For specimen S1, 11 selection points could be identified and were used for fitting of a first order (affine) polynomial. Thereafter, the \textbf{EBSD} data is aligned, warped, linearly interpolated to the same grid as that of the \textbf{Microstructure} data.

The results of this alignment step are displayed in \fig \ref{fig:fig5}b, where an SE/BSE overlay of \textbf{Microstructure} with aligned \textbf{EBSD} grain boundaries is shown. Overall, the affine displacement fields provide good alignment of the \textbf{EBSD} data to the \textbf{Microstructure} data. Alignment accuracy is measured at several distinct points, not used in the fitting routine, resulting in a misalignment below $100\ $nm (not shown).

Despite the large (non-linear) warping in the \textbf{EBSD} data, alignment to the \textbf{Microstructure} data using affine (first order) displacement fields was found to yield the best (global) alignment results for these specimens. Several tests were performed using higher order polynomials, which resulted in a marginal (and very localized) error reduction around the selection points, with larger misfits between the \textbf{EBSD} grain and \textbf{Microstructure} BSE boundaries away from the selection points, see the zoom figures in \fig \ref{fig:fig5}b. Based on these test results with few available homologous markers, the first order polynomial order works best, however, if many more homologous markers are available in the \textbf{EBSD} map, then second order warping may further improve the results.

\subsubsection{Alignment of the \textbf{Microstructure} and \textbf{SEM-DIC} to \textbf{Reference} Datasets}

\begin{figure}[H]
	\centering
    \includegraphics[width=0.6\textwidth]{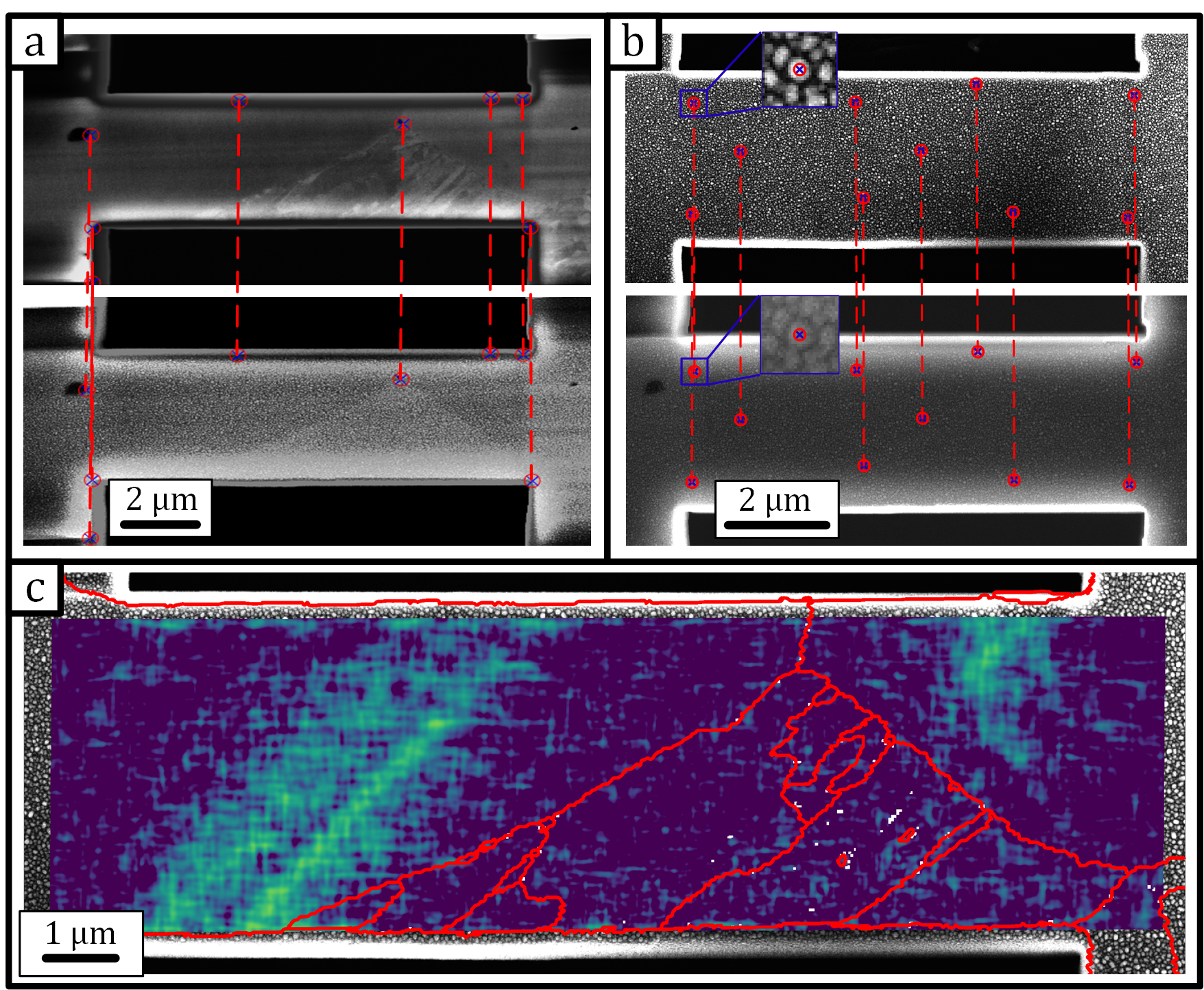}
    \caption{\small \textit{Alignment of \textbf{Microstructure} and \textbf{SEM-DIC} data sets to \textbf{Reference} data set, demonstrated for specimen S1. (a) Connected homologous points between \textbf{Microstructure} BSE/SE overlay and \textbf{Reference} IB-SE/BSE overlay data and (b) connected selection points between \textbf{SEM-DIC} IB-SE and \textbf{Reference} IB-SE, with insets illustrating how individual points are selected. (c) \textbf{SEM-DIC} to \textbf{Reference} alignment results, shown by plotting of the equivalent strain field with \textbf{EBSD} grain boundary overlay.}}
    \label{fig:fig6}
\end{figure}

Both the \textbf{Microstructure} and \textbf{SEM-DIC} data sets are aligned, warped and interpolated to the grid of the \textbf{Reference} dataset using the \textit{points-based alignment} module. \fig \ref{fig:fig6}a shows 8 connected homologous points (a combination of geometrical features, microstructural features and gauge topography) between an BSE/SE overlay of the \textbf{Microstructure} data and a BSE/IB-SE overlay of the \textbf{Reference} data. Additionally, individual InSn DIC speckles are employed as (10) homologous points between \textbf{SEM-DIC} (IB-SE) and IB-SE \textbf{Reference} data, as demonstrated in \fig \ref{fig:fig6}b. It can be observed that the displacement fields are described (almost purely) by RBM, as connection lines are closely parallel and of equal length for both combinations. An affine displacement field is therefore used for both alignments, as it is capable of correcting for RBM (including small rotations) and may also correct for any minor warping which might be present due to, for example, sample drift.

The results of the \textbf{SEM-DIC} to \textbf{Reference} alignment step can be seen in \fig \ref{fig:fig6}c, showing an overlay of the \textbf{EBSD} grain boundaries over the \textbf{SEM-DIC} equivalent strain data. Note that the \textbf{EBSD} grain boundaries are aligned to the \textbf{Reference} dataset using indirect alignment via \textbf{Microstructure}, therefore, the misalignment of \textbf{EBSD} grain boundaries to the \textbf{Microstructure} data is superimposed on the misalignment of the current step.

For \textbf{SEM-DIC} to \textbf{Reference} alignment, many more DIC speckles are available that could be used to enhance the alignment accuracy, which is estimated to be half a speckle or $15\ $nm. Alternatively, alignment could be performed through global DIC, since both datasets show the same speckle pattern, although at different brightness and contrast. Such an approach could result in sub-pixel ($<7\ $nm) alignment accuracy.

\subsubsection{Alignment of Rear to Front}
\begin{figure}[H]
	\centering
    \includegraphics[width=0.7\textwidth]{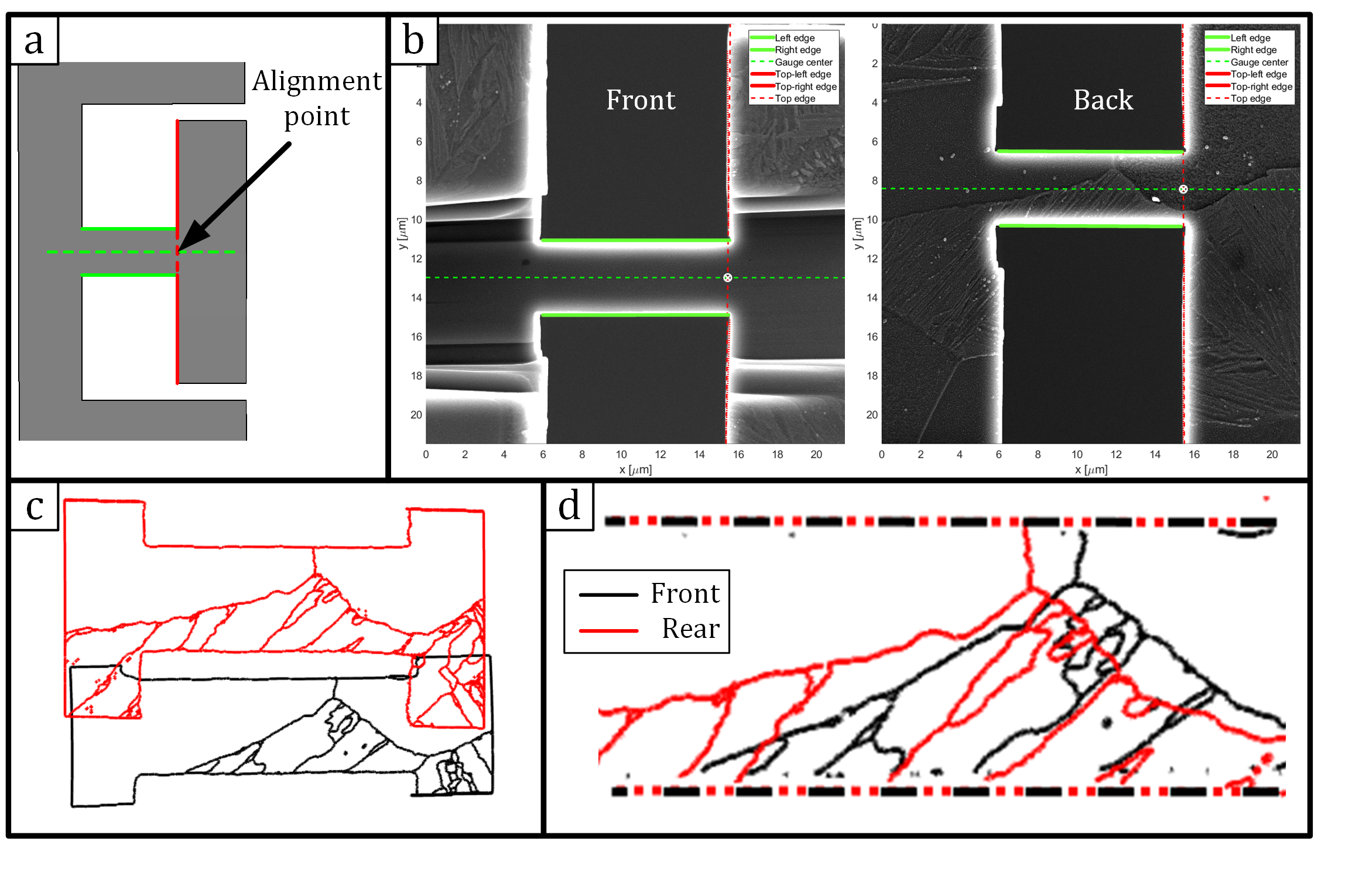}
    \caption{\small \textit{Alignment of rear to front \textbf{Reference} datasets, illustrated for specimen S1. (a) Schematic representation of edge-based alignment. The solid green lines indicate the two sides of the gauge section and are used to determine the gauge center line (dashed green line). The solid red lines indicate the gripper contact surface, from which the average contact surface (dashed red line) is derived. (b) SE images of the \textbf{Reference} dataset of both front and rear. The theoretically identical alignment point is indicated with the white marker. (c) Grain boundaries of front (black) and rear (red) data sets in one graph, visualizating the original misalignment between the datasets. (d) After alignment, zoom of the gauge with front (black) and rear (red) EBSD grain boundaries and the respective gauge edges (dashed lines), extracted from the \textbf{Reference} SE binary edge image.}}
    \label{fig:fig7}
\end{figure}

After all front and rear data sets are aligned in the, respective, front or rear \textbf{Reference} dataset configuration, the rear \textbf{Reference} data must be aligned to the front \textbf{Reference} data. Despite careful specimen selection and production, through-thickness changes in the microstructure will always be present, preventing correlation of the front and rear \textbf{Reference} datasets based solely on micro-structural features. Therefore, alignment is performed using the \textit{edge-based alignment} module. For nano-tensile specimens, \textit{edge-based alignment} relies on the edges of the T shaped specimen, which are visible on both sides of the specimen, to determine a theoretically identical alignment point (\fig \ref{fig:fig7}a). These edges are considered to be at the same in-plane position, as they were created in the same FIB milling step.

Canny edge detection is employed to create a (binary) edge image from the \textbf{Reference} SE image, which is used to identify the gauge edges (solid green lines) and the gripper contact surfaces (solid red lines) for the front and rear (\fig \ref{fig:fig7}b). The gauge edges and contact surfaces are averaged to determine the gauge center (dashed green lines) and average contact surface (dashed red lines), respectively. The theoretically identical alignment point, i.e. the intersection of the gauge center and gripper contact surface, is identified in both datasets (white marker) and is used to determine the translational misalignment between the front and rear \textbf{Reference} datasets. Rotational misalignment is determined using the relative angle between the two gauge center lines. The RBM displacement field necessary for alignment of the rear \textbf{Reference} dataset to the front \textbf{Reference} dataset is constructed by superimposing the translational and rotational misalignments. The rear \textbf{Reference} dataset, with all other rear datasets included, is aligned with the front \textbf{Reference} dataset by applying the RBM displacement fields resulting in the alignment shown in \fig \ref{fig:fig7}d.

\subsubsection{Alignment of \textbf{Post-Mortem} BSE to forward-deformed \textbf{Reference}}
\label{sec:forward}

Up to this point, all data used in the alignment framework was considered to be in the undeformed configuration, such that only translation, rotation or (minor) warping were required for alignment. However, \textbf{Post-Mortem} images, or \insitu images besides those used for DIC, may also require alignment to the microstructure and strain fields, e.g., to investigate the evolution of microstructure-resolved strain fields into cracks and/or damage \cite{Vermeij2021}. Therefore, as illustrated schematically in \fig \ref{fig:fig3} (bottom-right), the \textbf{Reference} dataset is first forward transformed to the deformed-configuration, updating its position field by adding the (aligned) DIC displacement field values to the position vector values (purple arrows in \fig \ref{fig:fig3}). Thereafter, the \textbf{Post-Mortem} image is aligned to this \textbf{Forward-Deformed Reference} using point-based alignment, such that all data can be plotted in the deformed configuration.

\section{Case Study: Incompatible Martensite-Ferrite Interface}
\label{sec:casestudy}

The capabilities and results of the full nanomechanical testing framework will be demonstrated here on specimen S2, which contains a single ferrite-martensite interface with high crystallographic misorientation along the full gauge-length. A full overview of all aligned data of specimen S2, in deformed and amplified-deformed configuration, will illustrate how the specimen deforms and how the plasticity occurs in the microstructure on two sides of the specimen. Subsequently, we employ a custom DIC code with varying DIC subset sizes, with and without "deconvolution" (explained below), on two smaller areas, employing different strain calculation methods to highlight the highest achievable spatial strain resolution and to identify both diffuse ferrite plasticity and discrete martensite plasticity. Finally, we utilize the aligned front\&rear-sided strain and microstructure data fields to interpret the 3D and through-thickness deformation behaviour. 

\subsection{Overall Microstructure and Deformation}
\begin{figure}[H]
	\centering
    \includegraphics[width=0.8\textwidth]{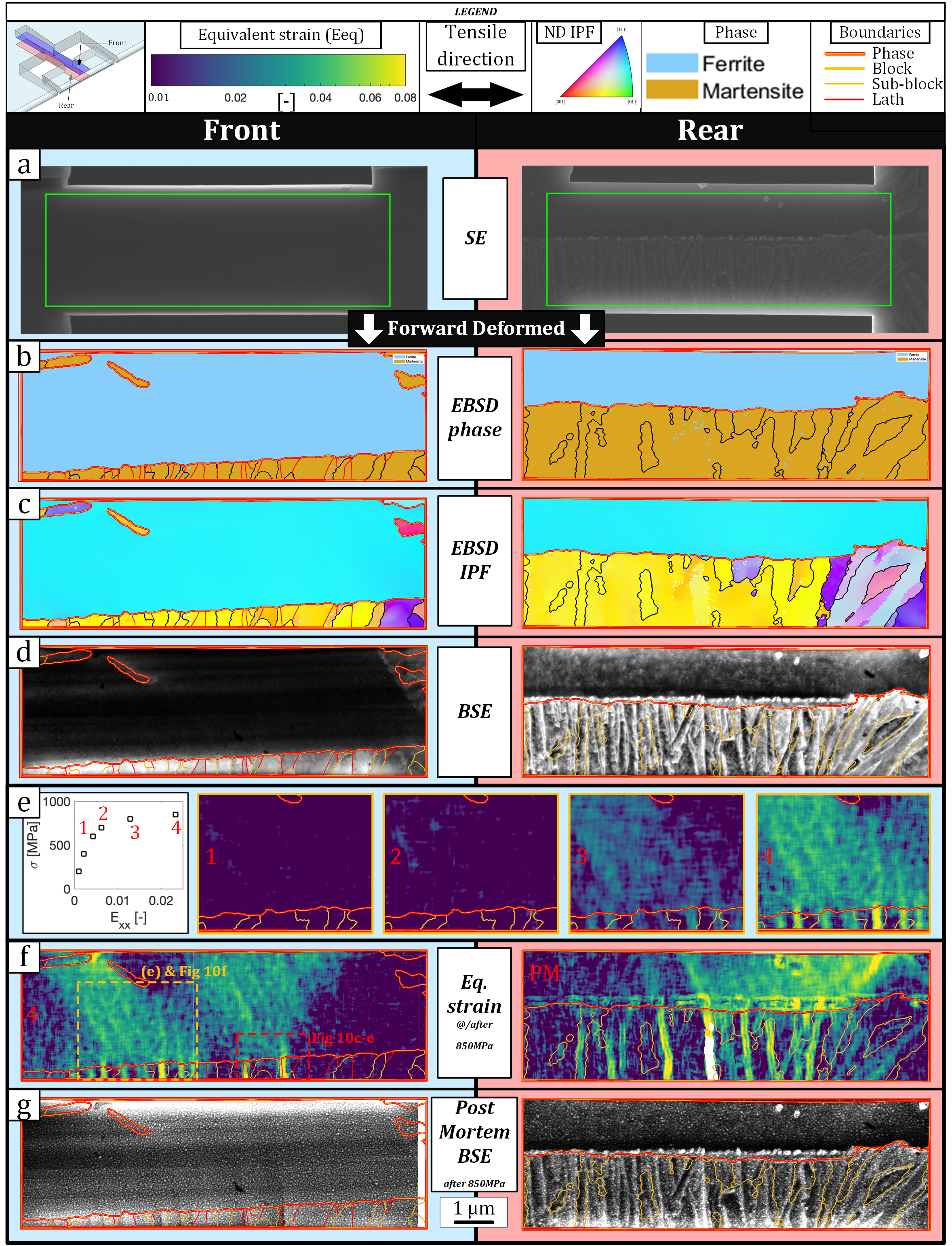}
    \caption{\small \textit{Overview of front (left, blue background) and rear (right, red background) aligned data of specimen S2 in the forward-deformed (except for (a)) configuration. (a) SE images of the gauge in undeformed configuration, with the green rectangle outlining the ROIs used for the rest of the subfigures, wherein DIC data is available, (b) EBSD phase map indicating the presence of ferrite or martensite, (c) EBSD ND IPF map, (d) BSE microstructure maps showing, through channeling contrast, fine microstructure morphology such as lath boundaries, which are manually drawn over the front datasets with red lines. Also includes EBSD grain and phase boundary overlay. (e) Global stress-strain curve with equivalent strain fields (on a small area on the front side marked in orange in (f)) at 4 increments as indicated. (f) Equivalent strain fields on the front (at 850 MPa) and the rear (unloaded after 850 MPa), with overlaid BSE-derived lath boundaries (front) and EBSD grain and phase boundaries (front and rear). The front strain field shows two smaller areas that are shown in more detail in (e) and in \fig \ref{fig:fig10}. (g) \textit{Post-mortem} high eV BSE images, with phase, grain and lath boundary overlay, showing signs of cracks and localizations at high strain positions at lath boundaries.}}
    \label{fig:fig8}

\end{figure}

 All the steps of the framework, as described in Section \ref{sec:framework}, are applied to specimen S2 and result in the aligned data sets as shown in \fig \ref{fig:fig8}, in which all plotting (except for the SE image in (a)) is done in the deformed configuration, which is hardly visible by eye due to the small global strain of $\sim 0.02$. Figures \ref{fig:fig8}b,c show the aligned EBSD phase map (martensite-ferrite identified by thresholding of the EmSphInx Confidence Index (CI)) and ND IPF map respectively, with an overlay of phase, grain and lath boundaries, as described in the legend. The specimen predominantly consists of a single through-thickness ferrite grain on the top side, with several martensite variants on the bottom side. The ferrite-martensite interface is straight and runs through-thickness, at an angle, from front to rear, resulting in more martensite on the rear than on the front. The location of the nano-tensile specimen on the wedge was carefully selected to contain the harder martensite along the full length of the specimen gauge, to assure that deformation crosses the phase boundary under global tension. According to theory, misorientations between laths are almost zero \cite{Morito2006a}, inhibiting easy identification with EBSD. This explains why in BSE channeling contrast imaging (\fig \ref{fig:fig8}d) clearly more boundaries are visible than the few boundaries identified by EBSD. These additional low-misorientation boundaries, which are most likely lath boundaries, are drawn by hand in red over the BSE image, in addition to the EBSD based grain and phase boundaries (\fig \ref{fig:fig8}d), and can also be used as an overlay for all other data sets since these are all aligned. The channeling contrast quality on the rear side is lower than that on the front, due to the topography caused by electropolishing (which was milled away by FIB on the front). Therefore, to prevent misidentification, lath boundaries are only drawn on the front side. 
 
The microstructure-correlated strain maps of front and rear in \fig \ref{fig:fig8}f, with overlay of all boundaries, indicate that plasticity is discrete in martensite and diffuse in ferrite, with a sharp transition at the ferrite-martensite interface. The plasticity protrusions from martensite into the ferrite would likely evolve into the so-called "near-interface damage" in bulk material, which has been studied in detail by Liu \textit{et al.} \cite{LeiLiu2020}. In martensite, slip bands clearly align with the lath and variant boundaries, which are in turn parallel to the martensite habit plane trace, as checked with a prior austenite grain reconstruction, discussed further in Section \ref{sec:frontRear}. In the ferrite away from the interface, plasticity spreads out into multiple directions and appears to be rather diffuse, inhibiting robust trace identification. In \fig \ref{fig:fig8}e, we also show the global stress-strain curve of the nano-tensile test (global strain calculated by averaging of the DIC strain data) and equivalent strain fields at several increments, as indicated in the stress-strain curve. In increment 2, plasticity initializes in the ferrite, which is expected since ferrite is the softer phase. Subsequently, in increment 3, plasticity in ferrite and localization in martensite occur concurrently, which is expected, considering that the ferrite and martensite need to deform roughly in parallel, which means strong plasticity on one side (e.g. the soft side) needs to be followed by the other side. The fact that the deformation pattern in both ferrite and martensite is very similar between increments 3 and 4, with only a difference in strain amplitude, underlines the robustness of the SEM-DIC strain measurement. To study the strain accuracy in more detail, two small near-interface areas are marked in \fig \ref{fig:fig8}f (orange and red dashed rectangles), which will be subject to a further high-resolution investigation in \fig \ref{fig:fig10}, with a focus on DIC parameters, deconvolution and strain calculation methods. The orange area is also used in \fig \ref{fig:fig8}e. Finally, the \textit{post-mortem} BSE images in \fig \ref{fig:fig8}g clearly shows several sharp localizations in the martensite on the front, which correspond exactly to the strongest strain bands in the martensite. On the rear, the localizations are not as clearly visible in the BSE contrast, due to the significant topography of the specimen, however, comparison with the strain map confirms that the slip bands closely align with the substructure boundaries. Moreover, these \textit{post-mortem} BSE images, which are carefully aligned to the forward-deformed microstructure and strain fields (see Section \ref{sec:forward}), are well suited for the identification of damage and to study any preceding plasticity mechanisms at the same position \cite{Vermeij2021}.
 
\begin{figure}[H]
	\centering
    \includegraphics[width=1\textwidth]{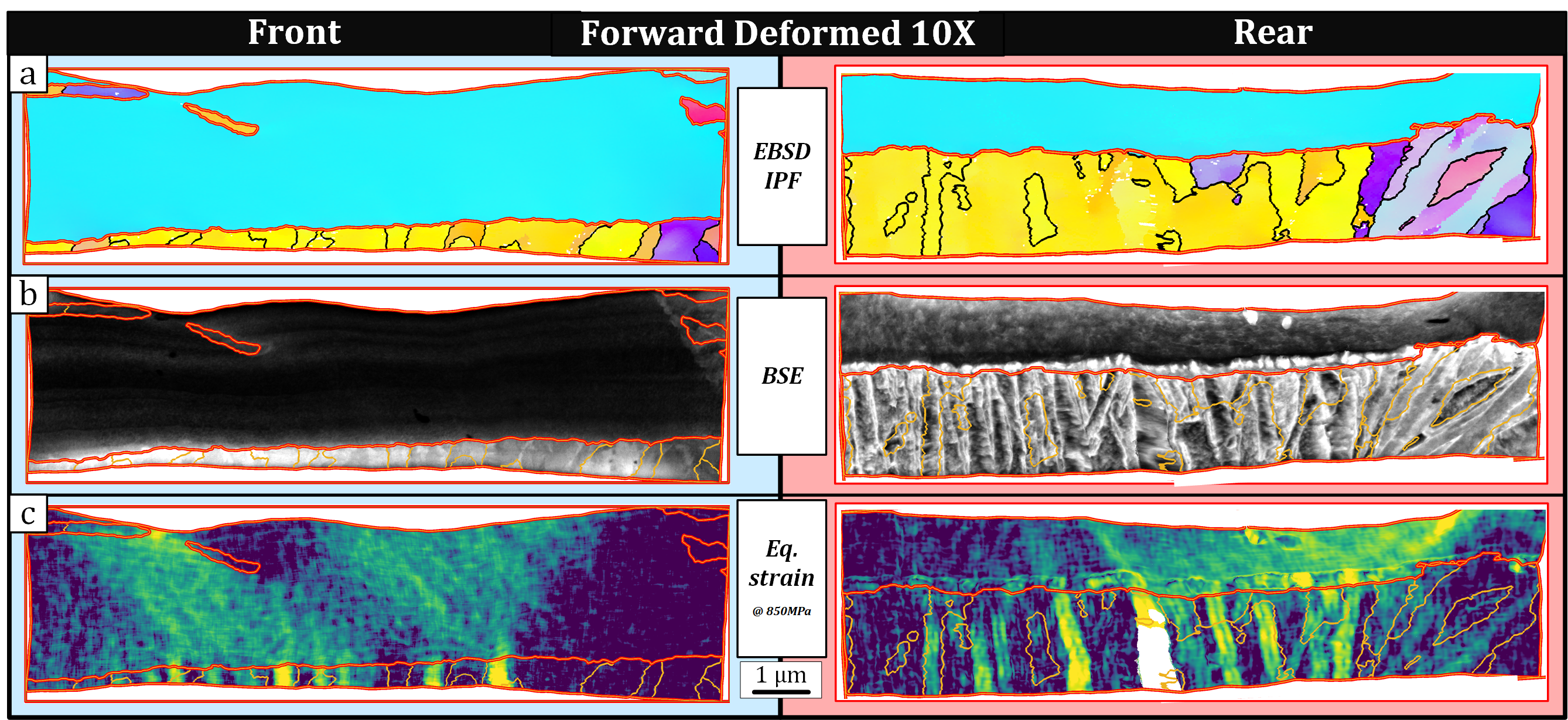}
    \caption{\small \textit{10X-amplified forward-deformed maps of front and rear of specimen S2, for visual interpretation of the deformation. The initial (undeformed) shape of the area of interest was the same rectangle as in \fig \ref{fig:fig8}. (a) EBSD ND IPF maps, (b) BSE microstructure maps and (c) equivalent strain maps. The legend of the maps is that of \fig\ref{fig:fig8}.}}
    \label{fig:fig9}
\end{figure}

The alignment of microstructure and DIC displacement data also allows plotting of all data in an "amplified" forward-deformed configuration, as illustrated in \fig \ref{fig:fig9}. This is achieved by multiplying the displacement field values with an "amplification factor", after which this amplified displacement field is added to the undeformed position field for plotting. An amplification factor of 10 was employed in \fig \ref{fig:fig9}a,b,c for plotting of IPF, BSE microstructure and equivalent strain field respectively, for front and rear, illustrating how the deformation occurs. This allows a direct comparison to simulations, wherein this "amplified" forward-deformed configuration is common practice. Some interesting observations include: (i) the smoothness of the ferrite edge versus the rather discrete steps on the martensite edge denote diffuse versus discrete plasticity; (ii) the necking behaviour, without any significant in-plane shear deformation, suggests partly out-of-plane deformation; and (iii) a "double" necking can be observed on the front side, likely caused by the small martensite islands on the top that partially blocks the ferrite plasticity near the edge.

\subsection{High-resolution DIC Study}

Next, we will highlight both the accurate microstructure-to-strain alignment and the high spatial strain resolution, which will be pushed to the limit through optimized DIC, "deconvolution" and strain calculation methods. In martensite, the exact location of the discrete plasticity needs to be determined, while in ferrite, we want to unravel the diffuse slip bands. Improvements of the spatial strain resolution in SEM-DIC may be achieved by: (i) decreasing the nano-DIC speckle pattern size, (ii) decreasing the subset size, (iii) employing the recently proposed method of "deconvolution" of the DIC displacement fields \cite{grediac2019robust}, (iv) optimizing strain calculation methods and/or (v) application of advanced DIC algorithms such as subset splitting \cite{poissant2010novel, BOURDIN2018307}. Although the InSn nano-DIC speckle pattern size \cite{Hoefnagels2019} could be decreased further for future experiments, we choose to focus here on points (ii), (iii) and (iv), while (v) subset splitting requires distinct slip steps that are clearly not present in ferrite. As such, we performed local DIC (see Table \ref{Table:DIC_params2} for all the parameters) on the front side using a varying subset size of 17, 29 and 41 pixels ($119\ $nm, $203\ $nm and $287\ $nm, respectively). Next, deconvolution was performed according to the approach proposed by Grediac \textit{et al.} \cite{grediac2019robust}. This deconvolution algorithm relies on the fact that DIC displacements correspond, when omitting the effect of the pattern \cite{sur2021biases}, to the physical displacement convoluted with a Savitzky-Golary kernel \cite{savitzky1964smoothing, schreier2002systematic}, which can be directly related to the subset size, shape and order. The deconvolution consists of correcting the higher order spatial derivative terms, which are computed by differentiation. These derivatives are calculated in the frequency domain, and a low-pass filter is introduced to avoid the amplification of high frequencies, e.g. arising from measurement noise, remaining SEM scanning artefacts, and potential pattern bias, with the threshold set to $f_{max} = 1/p_{min}$, wherein $p_{min}$ is the minimum wavelength of a signal that can be picked up by the deconvolution. The deconvolution is applied to the displacement fields resulting from the different subset sizes, using $p_{min}$ values of 25, 39 and 55 pixels for the data with subset size 17, 29 and 41 pixels, respectively. Additionally, we consider two strain calculation methods: (i) computation of the displacement gradient tensor $\vec{\nabla}_0\vec{u} $ through the direct (central differences) spatial gradient of $\vec{u}$ from both the ($\text{i}_\text{a}$) regular and ($\text{i}_\text{b}$) the deconvoluted displacement fields (after applying a gaussian filter with standard deviation of 3 pixels) and (ii) determination of the displacement gradient tensor $\vec{\nabla}_0\vec{u}$ from the correlated affine subset shape functions inside each subset (i.e. at each pixel position, since the step size was 1), which we call "Subset Internal Strains". No filtering is applied here. For both methods, Green-Lagrange and 2D equivalent strains are computed according to the equations given in Section \ref{sec:dicmethod}.

\begin{table}[H]\centering
\caption{\textit{\small Custom local DIC code correlation parameters}}
\label{Table:DIC_params2}
\ra{1.3}
\begin{tabular}{@{}lrl@{}}\toprule
\textbf{Parameter}  & \textbf{Value}    & \textbf{Unit} \\
\midrule
Software           & Home made                &            \\ 
Image filtering           & -                &           \\ 
Subset size           & 17 / 29 / 41                & pix           \\ 
Deconvolution: $p_min$           & 25 / 39 / 55                & pix           \\ 
Step size             & 1                 & pix           \\ 
Matching criterion & SSD & \\
Interpolant & Cubic Spline & \\
Strain window & 3x3  & pix \\ 
Virtual Strain Gauge size \cite{dicguide2018}  & 19 / 31 / 43   & pix\\ 
Subset Shape function        & Affine            &               \\ 
\bottomrule
\end{tabular}
\end{table}

\begin{figure}[H]
	\centering
    \includegraphics[width=0.75\textwidth]{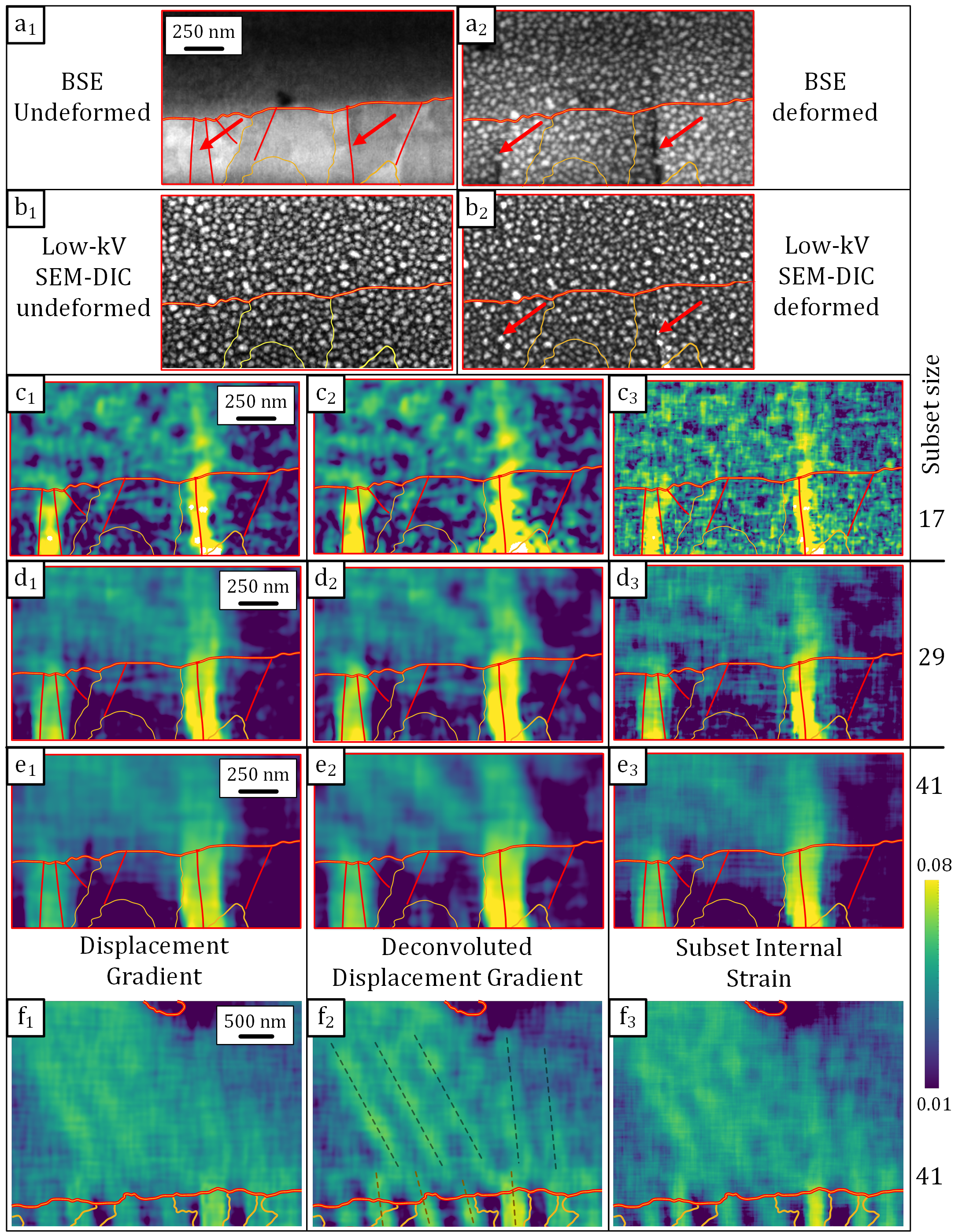}
    \caption{\small \textit{High-resolution investigation of the effect of strain calculation methods, DIC deconvolution and varying DIC subset sizes on small area 1 (a-e), which focuses on martensite deformation, and area 2 (f), focused on ferrite deformation. Both areas are indicated on the front strain field in \fig \ref{fig:fig8}f. For area 1, ($\text{a}_1$) shows undeformed BSE microstructure with overlay of phase, grain and lath boundaries, ($\text{a}_2$) \textit{post-mortem} high-eV BSE scan, showing strong localizations (local nano-damage) as indicated by the red  arrows, and ($\text{b}_{1,2}$) low-eV IB-SE images, as used for SEM-DIC correlation, in the ($\text{b}_1$) undeformed and ($\text{b}_2$) final deformed state. ($\text{c}_{1,2,3}$, $\text{d}_{1,2,3}$, $\text{e}_{1,2,3}$, $\text{f}_{1,2,3}$) equivalent strain fields of the two areas, computed with a subset size (in pixels) of (c) 17, (d) 29 and (e,f) 41. The columns show strain fields computed through ($\text{c}_{1}$, $\text{d}_{1}$, $\text{e}_{1}$, $\text{f}_{1}$) direct (nearest neighbour) gradient of the displacements, ($\text{c}_{2}$, $\text{d}_{2}$, $\text{e}_{2}$, $\text{f}_{3}$) direct (nearest neighbour) gradient of the deconvoluted displacements and ($\text{c}_{3}$, $\text{d}_{3}$, $\text{e}_{3}$, $\text{f}_{3}$) subset internal strain, for which the strain tensor is derived from the values of the degrees of freedom of the affine subset shape functions, for each subset (at each pixel) individually. In ($\text{f}_2$) we mark several clear slip traces in ferrite with black dashed lines and the martensite localizations that cross the interface with brown dashed lines. Note that the scale bar in (f) is different from that in the rest of the figure. All strain fields also contain overlays of phase, grain and lath boundaries.}}
    \label{fig:fig10}
\end{figure}

For the evaluation of the results of these correlations, we focus on two smaller areas on the front side (indicated in \fig \ref{fig:fig8}f), as shown in \fig \ref{fig:fig10}, which mainly comprises of an area with martensite activity in \fig \ref{fig:fig10}a,b,c,d,e, with a minor focus on ferrite in \fig \ref{fig:fig10}f. For the martensite area, \fig \ref{fig:fig10}a and \ref{fig:fig10}b show BSE images of the undeformed (without nano-DIC pattern) and deformed (with nano-DIC pattern) configuration respectively. Two strong localizations can be clearly observed (indicated by the red arrows) near and on top of a lath boundary. While clearly visible in the (high-eV) \textit{post-mortem} BSE image, the localizations are not so apparent in the SEM-DIC images (\fig \ref{fig:fig10}$\text{b}_1$,$\text{b}_2$), since IB-SE and low-eV imaging was performed such that only the nano-DIC pattern was imaged. Therefore, the localizations do not significantly hinder the DIC correlation, so that they can still be observed in the strain maps as sharp strain bands in \fig \ref{fig:fig10}c,d,e, where each row of subfigures shows strain fields computed with a certain subset size, and each column corresponds to one of the three strain calculation methods. The martensite strain localizations are clearly more narrow for smaller subset sizes, although the noise in the strain level increases at the same time. For this discrete plasticity, the deconvolution does not seem to improve the spatial resolution, likely due to the limiting factor of the $p_{min}$ value, which needed to be rather high here to avoid amplification of noise in the DIC results. However, the Subset Internal Strain fields show a clear improvement of spatial strain resolution as compared to the other strain calculation methods, which could be due to the fact that no filtering was applied. The advantage of the deconvolution however is quite clear for the small ferrite area in \fig \ref{fig:fig10}f, with the deconvoluted strain field showing the diffuse ferrite strain bands, likely due to plastic slip, most clearly, as indicated by the black dashed lines in \fig \ref{fig:fig10}$\text{f}_2$. This improvement for diffuse strains (as opposed to discrete strains) can likely be explained by the inherently lower spatial frequency of these diffuse bands, whereby the deconvolution is not inhibited by the $p_{min}$ (or equivalently $f_{max}$) values. Note that these clear strain bands, clarified through deconvolution, are only apparent in the strain data of a subset size of 41 pixels, likely because this provides a more stable correlation that is less sensitive to noise. In all, the analysis in \fig \ref{fig:fig10} shows that by selecting the most appropriate DIC strain calculation method and subset size, different features of the deformation can be exposed.

Next, we assess the microstructure-to-strain alignment accuracy by considering the localization which occurs exactly at the lath boundary (indicated by the red arrow in \fig \ref{fig:fig10} on the right side). In all strain maps of this area (\fig \ref{fig:fig10}c,d,e) the strain localization is centered on this martensite lath boundary, which proves that the microstructure-to-strain alignment is good enough to related specific plasticity events to features in the fine lath martensite microstructure. Indeed, in this case, we can likely classify this specific strain localization event as substructure boundary sliding, which has been shown to be an important plasticity mechanism in lath martensite \cite{du_plasticity_lath_martensite,Du2019} and dual-phase steel \cite{LeiLiu2020}. The localization event marked by the left red arrow appears to fall exactly between 2 lath boundaries, which is especially clear on the strain field with small subsets. Therefore, this strain localization appears to be either intra-lath slip along the habit plane \cite{ryou2020effect} or substructure boundary sliding over a lath boundary that is not clearly visible in the BSE images. This will be investigated in future work \cite{Vermeij2022boundarysliding}.

Finally, it is interesting to note that in the ferrite area in \fig \ref{fig:fig10}f the clear transition over the interface from discrete martensite plasticity to diffuse ferrite plasticity. While there is some connection between martensite and ferrite plasticity, as shown by the brown dashed lines at the bottom of \fig \ref{fig:fig10}f, these do not connect well to the ferrite slip bands, which is likely due to the plastic incompatibility between ferrite and martensite \cite{Vermeij2022dpcompatibility}.

\subsection{front\&rear-sided microstructure resolved strain fields}
\label{sec:frontRear}
\begin{figure}[H]
	\centering
    \includegraphics[width=0.7\textwidth]{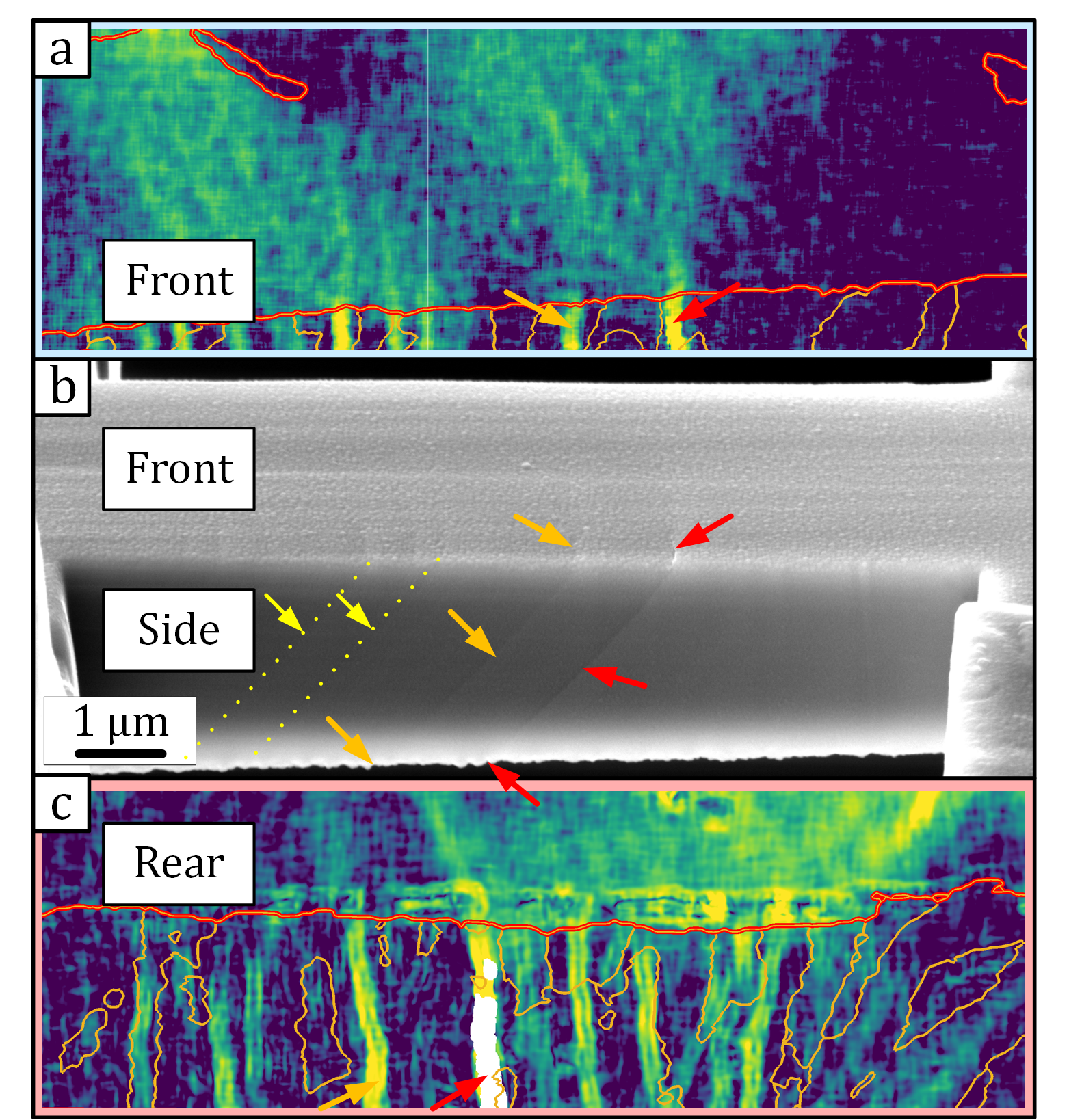}
    \caption{\small \textit{Interpretation of the 3D martensite deformation of specimen S2 by assessment of (a) front equivalent strain field, (b) \textit{post-mortem} SE side view and (c) rear equivalent strain field. The red and orange arrows both follow a martensite localization band that appears to be connected from the front, over the side, to the rear. The yellow dotted lines and arrows indicate two more slip traces on the side that are less visible. The strain field colorbar is equal to that in \fig \ref{fig:fig10}.}}
    \label{fig:fig11}
\end{figure}

Since both front and rear strain fields are available and are aligned with respect to each other and with respect to the microstructure, we can now attempt to link the deformations on both sides, with the aim of inferring how the deformation occurs in 3D. In \fig \ref{fig:fig11}a,c the front and rear strain fields are shown, while \fig \ref{fig:fig11}b contains a side-view of the specimen after deformation, in which several slip bands are clearly observed, as indicated by the arrows and dashed lines. Judging from the strain fields alone, one can already infer that the deformation mechanisms have the same character on front and rear: discrete (and vertical) in martensite and diffuse in ferrite. Moreover, the most prominent martensite strain localizations on front and rear can be followed over the side of the specimen, see the red and orange arrows, which means that these localizations run completely through the thickness, under an angle, as expected. On the ferrite side-view of the specimen, no discrete slip bands could be found (not shown), which agrees well with the large difference between front and rear for ferrite, which is also an indication that cross-slip is dominant in ferrite.

While the martensite slip bands align with the lath boundaries on front and rear, the lath boundaries are not visible on the side of the specimen. However, before deformation, we performed EBSD scans on a larger area around the specimen, which was used to perform a Prior Austenite Grain reconstruction \cite{hielscher2022variant} (not shown), through which the habit plane was determined. In lath martensite, lath boundaries are predominantly aligned with the habit plane, such that it can be used to estimate the 3D orientation of the lath boundary planes. For the current specimen, the habit plane aligns well with the martensite slip traces on the front, side and rear surfaces. This is a strong indication of martensite plasticity along the laths and/or their boundaries. For further verification of the 3D lath boundary configuration, specimen characterization of future specimens could be extended to include BSE and EBSD imaging of the sides of the specimens.

\section{Conclusions}
\label{sec:conclusions}
Advancement in the understanding of complex nanoscale multiphase interface and grain boundary mechanics requires measurement of nanoscale deformation mechanisms on targeted microstructure configurations, a challenge that requires combination of two state-of-the-art methods: (i) well-defined micro-deformations tests of carefully chosen specimens and (ii) measurement of nanoscale resolution microstructure-resolved strain fields. In this work, we presented a nanomechanical testing framework that addresses this challenge by integrating the following state-of-the-art testing and characterization methods:

\begin{itemize}
\item (I) specimen selection and fabrication: fabrication of micron-sized "1D" specimens that were isolated from the bulk microstructure at specific regions of interest;
\item (II) characterization and nano-DIC patterning: front\&rear-sided microstructure characterization and front\&rear-sided application of a recently developed ultrafine nano-DIC speckle pattern on the micro-specimens;
\item (III) nanoscale testing and DIC: front\&rear-sided high-resolution SEM-DIC strain mapping, under uniaxial loading conditions, aided by SEM scanning artefact correction and DIC deconvolution correction;
\item (IV) data alignment: alignment of all microstructure and strain data using a novel data alignment framework.
\end{itemize}

A case study on a particularly interesting type of dual-phase steel specimen with an incompatible (i.e. high crystallographically misoriented) ferrite-martensite interface showed how the very high spatial strain resolution (after optimization of DIC and application of deconvolution) and the careful microstructure-to-strain alignment gives insight into the martensite, ferrite and (near-)interface deformation mechanisms at the nanoscale. Moreover, the front\&rear-sided and aligned strain fields, in combination with the well-defined and isolated state of the multiphase specimen, gives opportunities to unravel the 3D deformations while only having access to in-plane strain fields. Additionally, the high degree of alignment between microstructure and strain fields allow plotting of microstructure and strain fields into an "amplified" deformed configuration, which eases interpretation of the general deformation of the specimen.

The ferrite-martensite interface specimen presented in this work shows how discrete martensite plasticity protrudes through the interface into the ferrite and transits further in the ferrite into diffuse plasticity. Through analysis of the SEM-DIC nanoscale strain fields, EBSD-based habit plane identification in the martensite and multi-sided observation, all aligned with the here-proposed data alignment framework, the martensite plasticity is identified to be aligned with the lath boundaries, confirming previous observations of soft habit plane martensite plasticity.

The isolated nature of these tests, under well-defined loading conditions, combined with front\&rear-sided full-field knowledge of microstructure-resolved displacement and strain fields, provides ample opportunities for more extensive and more quantitative comparison to complex multiphase numerical simulations than was possible so far. Yet, even without simulation, identification of the slip systems in complex phases such as martensite and bainite, and identification of nanoscale interface deformation mechanisms, is feasible after application of the here-presented nanomechanical testing framework.

\section*{Acknowledgements}
The authors acknowledge Marc van Maris, Chaowei Du, Lei Liu, Marc Geers and Ron Peerlings for discussions and (experimental) support. This research was carried out under project number S17012b in the framework of the Partnership Program of the Materials innovation institute M2i (www.m2i.nl) and the Technology Foundation TTW (www.stw.nl), which is part of the Netherlands Organization for Scientific Research (http://www.nwo.nl). B. Blaysat is grateful to the French National Research Agency (ANR) and to the French government research program "Investissements d'Avenir" for their financial support (ICAReS project, N ANR-18-CE08-0028-01 \& IDEX-ISITE initiative 16-IDEX-0001, CAP 20-25).

\section*{Declarations}
The authors state that regarding the research and writing of the manuscript, there are no conflicts of interest and that all the authors consent to the content of the manuscript.

\section*{Code \& Data availability}
The Matlab code for the full data alignment framework, together with a small example, will be available on GitHub (\url{https://github.com/Tijmenvermeij/NanoMech_Alignment_Matlab}). The full dataset is available upon request.

\centering
\noindent\rule{8cm}{0.4pt}

\clearpage
%
\bibliography{lib}

\end{document}